\documentstyle[12pt]{article}

\begin{document}

\noindent {\small CITUSC/00-014\hfill \hfill hep-th/0003100 \newline
}

{\small \hfill }

{\vskip0.5cm}

\begin{center}
{\Large {\bf Two-Time Physics in Field Theory}}{\footnote{%
This research was partially supported by the US. Department of Energy under
grant number DE-FG03-84ER40168.}}{\Large {\bf \ }}

\bigskip

{\vskip0.5cm}

{\bf Itzhak Bars}

{\vskip0.5cm}

CIT-USC Center for Theoretical Physics

and

{Department of Physics and Astronomy}

{University of Southern California}

{\ Los Angeles, CA 90089-2535, USA}

{\vskip0.5cm}

{\bf Abstract}
\end{center}

A field theory formulation of two-time physics in $d+2$ dimensions is
obtained from the covariant quantization of the constraint system associated
with the OSp$\left( n|2\right) $ worldline gauge symmetries of two-time
physics. Interactions among fields can then be included consistently with
the underlying gauge symmetries. Through this process a relation between
Dirac's work in 1936 on conformal symmetry in field theory and the more
recent worldline formulation of two-time physics is established while
providing a worldline gauge symmetry basis for the field equations in $d+2$
dimensions. It is shown that the field theory formalism goes well beyond
Dirac's goal of linearizing conformal symmetry. In accord with recent
results in the worldline approach of two-time physics, the $d+2$ field
theory can be brought down to diverse $d$ dimensional field theories by
solving the subset of field equations that correspond to the ``kinematic''
constraints. This process embeds the one ``time'' in $d$-dimensions in
different ways inside the $d+2$ dimensional spacetime. Thus, the two-time $%
d+2$ field theory appears as a more fundamental theory from which many
one-time $d$ dimensional field theories are derived. It is suggested that
the hidden symmetries and relations among computed quantities in certain $d$%
-dimensional interacting field theories can be taken as the evidence for the
presence of a higher unifying structure in a $d+2$ dimensional spacetime.
These phenomena have similarities with ideas such as dualities, AdS-CFT
correspondence and holography.\newpage

\section{Introduction}

In 1936 Dirac invented a field theory approach for rewriting conformal field
theory in four dimensions in a manifestly SO$\left( 4,2\right) $ covariant
form in six dimensions \cite{Dirac}. Dirac's fields $\Phi \left( X\right) $
depend on $6$ coordinates $X^{M}$ which have two timelike dimensions, just
like the dynamical coordinates $X^{M}\left( \tau ,\cdots \right) $ used in
the formalism of two-time physics on the worldline or worldvolume \cite{conf}%
-\cite{emgrav}. In the notation of \cite{conf}-\cite{emgrav} to label $%
X^{M}, $ with $M=+^{\prime },-^{\prime },0,1,2,3,$ Dirac's choice of
coordinates are as follows: Minkowski space coordinates $x^{\mu }$ are the
homogeneous coordinates $x^{\mu }=X^{\mu }/X^{+^{\prime }},$ with $\mu
=0,1,2,3$, while the extra coordinate $X^{-^{\prime }}$ is eliminated
through the SO$\left( 4,2\right) $ invariant constraint $X\cdot
X=-2X^{+^{\prime }}X^{-^{\prime }}+X_{\mu }X^{\mu }=0$. The extra
coordinates $X^{0^{\prime }},X^{1^{\prime }}$ given by $X^{\pm ^{\prime
}}=\left( X^{0^{\prime }}\pm X^{1^{\prime }}\right) /\sqrt{2}$ describe one
extra timelike and one extra spacelike dimensions. Dirac showed that the
free field equations for scalar, fermion and vector fields in $4$ dimensions 
$\phi \left( x^{\mu }\right) $ can be rewritten SO$\left( 4,2\right) $
covariantly in terms of fields $\Phi \left( X^{M}\right) $ that depend on
the $6$ coordinates, provided these fields also satisfy additional SO$\left(
4,2\right) $ covariant subsidiary conditions. Several authors pursued
Dirac's idea and extended it to interacting conformal field theories,
including conformally invariant Yang-Mills theories \cite{Kastrup}\cite
{salam}, but then Dirac's idea was forgotten for a long time. Recently this
approach has been applied to conformal gravity and its interactions with
conformal matter \cite{vasilev}.

Dirac's goal was to realize conformal symmetry linearly in 4+2 dimensional
field theory, and this remained the primary motivation for the work in the
literature that followed his paper. The goals and results of two-time
physics lie in more general directions, although conformal symmetry is
included as a special outcome in a particular gauge. In two-time physics
there is an underlying new gauge principle that is responsible for recasting
the $d+2$ dimensional theory as many possible $d$-dimensional theories. The
purpose of the present paper is twofold. First, to establish the
relationship between the gauge principles in two time physics on the
worldline and Dirac's approach in field theory; second, to demonstrate
directly in field theory that diverse one-time field theories emerge in $d$
dimensions from the same field equations in $d+2$ dimensions. It will be
seen that the path of derivation of $d$ dimensional field theories is in
precise correspondence with making gauge choices in the worldline theory,
the important step being the embedding of the time coordinate in $d$
dimensions in various ways inside the $d+2$ dimensions. In this way one can
see that the $d+2$ dimensional two-time theory plays a unifying role in a
new sense, including interactions.

Two-time physics in $d+2$ dimensions was developed independently in the
worldline (and worldvolume) formulation \cite{conf}-\cite{emgrav}, unaware
of the field theory formalism invented by Dirac which had been long
forgotten \footnote{%
I thank Vasilev for bringing to my attention his recent work, and informing
me of Dirac's work and the line of research that followed the same trend of
thought in relation to conformal symmetry.}. It was perhaps lucky that
ignorance of Dirac's approach permitted the free exploration and development
of new insights in the worldline formulation that were not necessarily
connected with conformal symmetry. Historically, the motivation for two-time
physics came from duality, and signals for two-timelike dimensions in
M-theory and its extended superalgebra including D-branes \cite{duff}-\cite
{nishino}. In particular certain dynamical attempts \cite{ibkounnas}\cite
{multisuper} to try to understand these phenomena directly paved the way to
the formalism in \cite{conf}. Two-time physics introduced a new gauge
principle - Sp$\left( 2,R\right) $ in phase space, and its generalizations -
that insures unitarity, causality and absence of ghosts. This takes care of
problems that naively would have arisen in a spacetime with two-timelike
dimensions. Morally speaking, this gauge symmetry is related to duality in a
generalized sense. The new phenomenon in two-time physics is that this gauge
symmetry can be used to obtain various one-time dynamical systems in $d$
dimensions from the same two-time action in $d+2$ dimensions, through gauge
fixing, thus uncovering a new layer of unification through higher
dimensions. In this paper we will show that the same insights can be
expressed in the language of field theory.

First we will show that the Sp$\left( 2,R\right) $ gauge symmetry (or OSp$%
\left( n|2\right) $ for spinning particles) provides a fundamental gauge
symmetry basis for Dirac's field equations in $d+2$ dimensions. In effect,
the field equations amount to imposing the non-Abelian OSp$\left( n|2\right) 
$ constraints in an SO$\left( d,2\right) $ covariant quantization of the
worldline-two-time physics theory, while the fields represent the gauge
invariant states. After reaching a two-time field theory formalism for
scalars, spinors, vectors and higher spin fields, field interactions
consistent with the underlying worldline gauge invariance is included. In
particular, interactions that are local in $d+2$ spacetime, such as
Yang-Mills or general reparametrizations, must satisfy certain ``kinematic''
field equations beyond the dynamical field equations, that are in complete
agreement with recent results obtained through background field methods in
two-time physics on the worldline \cite{emgrav}. The interacting field
theory constructed in this way is in agreement with the latest developments
in the Dirac approach included in \cite{vasilev}.

Second, it is shown that, depending on the path of coming down from $d+2$
dimensions to some chosen subset of $d$ dimensions, by solving the
``kinematic'' subset of the field equations, {\it the physical meaning of
the one-time field theory, as interpreted by an observer in the remaining }$%
d $ {\it dimensions, can be quite different}. In particular the natural SO$%
\left( d,2\right) $ Lorentz symmetry of the original field equations (in the
case of flat d+2 dimensional spacetime) can be interpreted in different ways
depending on the choice of the remaining $d$ coordinates. The resulting
one-time field theory has conformal symmetry if one follows Dirac's path
from $d+2$ to $d$, but with various embeddings of $d$ dimensions in $d+2$
dimensions one arrives at various one-time field theories. In the flat case,
all resulting $d$ dimensional field theories have new hidden SO$\left(
d,2\right) $ symmetries which are not necessarily conformal symmetries. Thus
the two-time field theory approach unifies classes of one-time physical
systems in $d$ dimensions that previously would have been thought of as
being described by $d$-dimensional field theories unrelated to each other.

Solving the ``kinematic'' subset of field equations amounts to a gauge
choice in the worldline formalism of two-time physics, and therefore the
physical interpretation of the remaining field theory agrees with similar
recent results in the worldline approach. The main essential new point
achieved through field theory is the inclusion of interactions in this new
type of unification.

These results hold at the level of classical field theory, which could be
thought of as the first quantization of the worldline theory. To extend them
to second quantized field theory (and analyze issues such as anomalies,
etc.) certain open problems in the field theoretic formulation of two-time
physics need to be understood. These may involve non-commutative geometry,
and they are briefly discussed in the last section. Analogies and
connections with other concepts in the literature, such as duality, AdS-CFT
and holography are also pointed out in the last section.

\section{Local and global symmetry}

The two-time worldline description of particle dynamics, in the absence of
background fields (i.e. ``free'' case\footnote{%
Although interactions are not explicitly present in the ``free'' action in $%
d+2$ dimensions, the solution of the constraints generates a class of
dynamics for the remaining degrees of freedom in $d$ dimensions after a
gauge is fixed. When background fields are present all possible particle
dynamics in $d$ dimensions (rather than only a class) can be described from
the point of view of two-time physics in $d+2$ dimensions, as shown in \cite
{emgrav}. We also mention that another generalization is space-time
supersymmetry, including a generalized local kappa supersymmetry \cite
{super2t}\cite{liftM}\cite{toyM}. This enriches both the local symmetries as
well as the global symmetries. The formalism has also been generalized to
strings and branes with limited success so far \cite{string2t} (although
full success is expected).}), is given by the $Sp(2,R)$ gauge theory
described by the action \cite{conf} 
\begin{eqnarray}
S_{0} &=&\frac{1}{2}\int d\tau \,D_{\tau }X_{i}^{M}X_{j}^{N}\varepsilon
^{ij}\eta _{MN}\equiv \int d\tau (\partial _{\tau }X_{1}^{M}X_{2}^{N}-\frac{1%
}{2}A^{ij}X_{i}^{M}X_{j}^{N})\eta _{MN}  \label{action} \\
&=&\int d\tau \left( \partial _{\tau }X^{M}P^{N}-\frac{1}{2}A^{11}X^{M}X^{N}-%
\frac{1}{2}A^{22}P^{M}P^{N}-A^{12}X^{M}P^{N}\right) \eta _{MN}.
\label{second}
\end{eqnarray}
Here $X_{i}^{M}(\tau )$ is an $Sp(2,R)$ doublet, consisting of the ordinary
coordinate and its conjugate momentum ($X_{1}^{M}\equiv X^{M}$ and $%
X_{2}^{M}\equiv P^{M}=\partial S_{0}/\partial X_{1M}$). The indices $i,j=1,2$
denote the doublet $Sp(2,R)$, they are raised and lowered by the
antisymmetric Levi-Civita symbol $\varepsilon _{ij}$. The gauge covariant
derivative $D_{\tau }X_{i}^{M}$ that appears in (\ref{action}) is defined as 
\begin{equation}
D_{\tau }X_{i}^{M}=\partial _{\tau }X_{i}^{M}-\varepsilon
_{ik}A^{kl}X_{l}^{M}.
\end{equation}
The local $Sp(2,R)$ acts as $\delta X_{i}^{M}=\varepsilon _{ik}\omega
^{kl}X_{l}^{M}$ and $\delta A^{ij}=\omega ^{ik}\varepsilon
_{kl}A^{lj}+\omega ^{jk}\varepsilon _{kl}A^{il}+\partial _{\tau }\omega
^{ij} $, where $\omega ^{ij}\left( \tau \right) $ is a symmetric matrix
containing the three Sp$\left( 2,R\right) $ gauge parameters and $A^{ij}$ is
the gauge field on the worldline. The second form of the action (\ref{second}%
) is obtained after an integration by parts so that only $X_{1}^{M}$ appears
with derivatives. This allows the identification of $X,P$ by the canonical
procedure, as indicated in the third form of the action.

The gauge fields $A^{11}$, $A^{12}=A^{21}$, and $A^{22}$ act as Lagrange
multipliers for the following three first class constraints that form the Sp$%
\left( 2,R\right) $ algebra 
\begin{equation}
X_{i}\cdot X_{j}=0\,\,\rightarrow \,\,\,X^{2}=P^{2}=X\cdot P=0,
\end{equation}
as implied by the local $Sp(2,R)$ invariance. It is precisely the solution
of these constraints that require that the global metric $\eta _{MN}$ has a
signature with two-time like dimensions. Thus, $\eta _{MN}$ stands for the
flat metric on a ($d,2$) dimensional space-time, which is the only signature
consistent with the equations of motion for the $Sp(2,R)$ gauge field $%
A^{kl} $, leading to non-trivial dynamics that can be consistently
quantized. Hence the global two-time $SO(d,2)$ is implied by the local $%
Sp(2,R)$ symmetry.

The explicit global $SO(d,2)$ invariance has the Lorentz generators 
\begin{equation}
L^{MN}=X^{M}P^{N}-X^{N}P^{M}=\varepsilon ^{ij}X_{i}^{M}X_{j}^{N}  \label{lmn}
\end{equation}
that are manifestly Sp$\left( 2,R\right) $ gauge invariant. As mentioned
above, different gauge choices lead to different particle dynamics in $d$
dimensions (relativistic massless and massive particles, non-relativistic
massive particle, H-atom, harmonic oscillator, particle in AdS$_{d-k}\times $%
S$^{k}$ background etc.) all of which have $SO(d,2)$ invariant actions that
are directly obtained from (\ref{action}) by gauge fixing. Since the action (%
\ref{action}) and the generators $L^{MN}$ (\ref{lmn}) are gauge invariant,
the global symmetry SO$\left( d,2\right) $ is not lost by gauge fixing. This
explains why one should expect a hidden (previously unnoticed) global
symmetry SO$\left( d,2\right) $ for each of the systems that result by gauge
fixing \cite{lifting}.

To describe spinning particles, worldline fermions $\psi _{a}^{M}\left( \tau
\right) $, with $a=1,2,\cdots ,n$ are introduced. Together with $X^{M},P^{M}$%
, they form the fundamental representation $\left( \psi
_{a}^{M},X^{M},P^{M}\right) $ of OSp$\left( n|2\right) $. Gauging this
supergroup \cite{spin} instead of Sp$\left( 2,R\right) $ produces a
Lagrangian that has $n$ local supercharges plus $n$ local conformal
supercharges on the worldline, in addition to local Sp$\left( 2,R\right) $
and local SO$\left( n\right) $. The full set of first class constraints that
correspond to the generators of these gauge symmetries are, at the classical
level, 
\begin{equation}
X\cdot X=P\cdot P=X\cdot P=X\cdot \psi _{a}=P\cdot \psi _{a}=\psi _{\lbrack
a}\cdot \psi _{b]}=0.  \label{ospn2gen}
\end{equation}
To have non-trivial classical solutions of these constraints (with angular
momentum) at least two timelike dimensions are required. The OSp$\left(
n|2\right) $ gauge symmetry can remove the ghosts of no more than two
timelike dimensions. Therefore, as in the spinless case, the signature is
fixed and the global symmetry of the theory is SO$\left( d,2\right) $. It is
applied to the label $M$ in $\left( \psi _{a}^{M},X^{M},P^{M}\right) $. The
global SO$\left( d,2\right) $ generators $J^{MN}$ that commute with all the
OSp$\left( n|2\right) $ gauge generators (\ref{ospn2gen}) are 
\begin{equation}
J^{MN}=L^{MN}+S^{MN}\,,\quad S^{MN}=\frac{1}{2i}\left( \psi _{a}^{M}\psi
_{a}^{N}-\psi _{a}^{N}\psi _{a}^{M}\right) .
\end{equation}

In this paper we will be interested in the covariant quantization of the
theory in a manifestly SO$\left( d,2\right) $ covariant formalism. This will
be used in the next section to construct the $d+2$ dimensional field theory.
The commutation rules are 
\begin{equation}
\left[ X^{M},P^{N}\right] =i\eta ^{MN},\quad \left\{ \psi _{a}^{M},\psi
_{b}^{N}\right\} =\,\eta ^{MN}\delta _{ab},
\end{equation}
while all other commutators among the basic degrees of freedom are zero. The 
$Sp(2,R)$ or OSp$\left( n|2\right) $ gauge constraints applied on the
Hilbert space are just enough to remove all negative-norm states
(``ghosts'') introduced by the two timelike dimensions \cite{conf}\cite{spin}%
, resulting in a unitary quantum theory. We will treat spinless particles as
a special case of OSp$\left( n|2\right) $ with $n=0,$ so we will state the
covariant quantization procedure directly for OSp$\left( n|2\right) .$

Since the constraints form a non-Abelian algebra one must choose a commuting
subset of operators to label the Hilbert space. In particular the local OSp$%
\left( n|2\right) $ labels and the global SO$\left( d,2\right) $ labels
correspond to simultaneously diagonalizable operators that include the
Casimir operators of both groups 
\begin{equation}
|OSp\left( n|2\right) \,\,labels;\,SO\left( d,2\right) \,\,labels>
\label{labels}
\end{equation}
The OSp$\left( n|2\right) $ quadratic Casimir operator that commutes with
all the generators in (\ref{ospn2gen}) is (before they are set to zero) 
\begin{eqnarray}
C_{2}\left( OSp\left( n|2\right) \right) &=&\frac{1}{8}\left(
X^{2}P^{2}+P^{2}X^{2}\right) -\frac{1}{16}\left( X\cdot P+P\cdot X\right)
^{2}  \label{c2ospn2} \\
&&+\frac{1}{4i}\left( X\cdot \psi _{a}P\cdot \psi _{a}-P\cdot \psi
_{a}X\cdot \psi _{a}\right) \\
&&+\frac{1}{32}\left( \psi _{\lbrack a}\cdot \psi _{b]}\right) \left( \psi
_{\lbrack a}\cdot \psi _{b]}\right) .
\end{eqnarray}
On the other hand, the global SO$\left( d,2\right) $ quadratic Casimir
operator is given by (orders of operators respected) 
\begin{eqnarray}
C_{2}\left( SO\left( d,2\right) \right) &=&\frac{1}{2}J^{MN}J_{MN}=\frac{1}{2%
}L^{MN}L_{MN}+\frac{1}{2}S^{MN}S_{MN}+L^{MN}S_{MN}\,,  \label{C2sod2} \\
\frac{1}{2}L^{MN}L_{MN} &=&\frac{1}{2}\left( X^{2}P^{2}+P^{2}X^{2}\right) -%
\frac{1}{4}\left( X\cdot P+P\cdot X\right) ^{2}+1-\frac{d^{2}}{4}\,, \\
\frac{1}{2}S^{MN}S_{MN} &=&\frac{1}{8}\left( \psi _{\lbrack a}\cdot \psi
_{b]}\right) \left( \psi _{\lbrack a}\cdot \psi _{b]}\right) +\frac{1}{8}%
n\left( d+2\right) \left( d+n\right) \,,  \label{ss} \\
L^{MN}S_{MN} &=&-i\left( X\cdot \psi _{a}P\cdot \psi _{a}-P\cdot \psi
_{a}X\cdot \psi _{a}\right) -\frac{1}{2}n\left( d+2\right) \,.
\end{eqnarray}
The extra constants arise from the re-ordering of quantum operators. In the
last two lines we have used $\psi _{a}\cdot \psi _{a}=n\left( d+2\right) /2$
that follows from the quantum relation. We see that the Casimir operator of
SO$\left( d,2\right) $ is related to the Casimir operator of OSp$\left(
n|2\right) $%
\begin{equation}
C_{2}\left( SO\left( d,2\right) \right) =4C_{2}\left( OSp\left( n|2\right)
\right) +\frac{1}{8}\left( d+2\right) \left( n-2\right) \left( d+n-2\right) .
\end{equation}
Similarly, higher Casimir operators of $SO\left( d,2\right) $ are also
related to Casimir operators of OSp$\left( n|2\right) $ except for ordering
constants.

One must demand that the physical states be singlets under the gauge
symmetry OSp$\left( n|2\right) $. This requires vanishing Casimir operators
of the gauge group, in particular $C_{2}\left( OSp\left( n|2\right) \right)
=0$. This leads to definite and unique eigenvalues for the SO$\left(
d,2\right) $ Casimir operators for physical states. Thus, on physical states
the quadratic Casimir operator must have the eigenvalue 
\begin{equation}
C_{2}\left( SO\left( d,2\right) \right) =\frac{1}{8}\left( n-2\right) \left(
d+2\right) \left( d+n-2\right) .  \label{Cospn2}
\end{equation}
Similarly, the higher Casimir eigenvalues for SO$\left( d,2\right) $ are
also fixed. Therefore, for given $d,n$ one must take a specific SO$\left(
d,2\right) $ representation to guarantee an OSp$\left( n|2\right) $ gauge
singlet. For example, for spinless particles ($n=0$) the quadratic Casimir
is fixed to $C_{2}=1-d^{2}/4$ (in the absence of background fields).

When the quantization is performed in a fixed gauge the same eigenvalue of
the Casimir operators must emerge for the dynamics of the remaining
dynamical system in $d$ dimensions for a fixed $n$. Indeed after careful
ordering of non-linear products of operators this is verified explicitly
(see \cite{conf}\cite{spin}\cite{lifting} for examples of non-covariant
quantization in several fixed gauges). The covariant quantization explains
why seemingly unrelated dynamics in $d$ dimensions (such as massless
relativistic particle, H-atom, harmonic oscillator in one less dimension,
particle in AdS$_{d-k}\times $S$^{k}$ for all $k,$ etc.) all must realize 
{\it the same unitary representation} of SO$\left( d,2\right) ,$ as they
indeed do.

\section{Fields, ``kinematics'' and ``dynamics''}

If the system is quantized in a fixed gauge, one time and one space
dimensions are eliminated, making the absence of ghosts and the one-time
nature of the system quite evident \cite{conf}\cite{lifting}. The quantum
theory is then expressed in terms of a wave equation in $d$ dimensions for
each one of the fixed gauges (e.g. for $n=0$ spinless particles:
Klein-Gordon, non-relativistic Schr\"{o}dinger, H-atom wave equation,
Klein-Gordon in AdS$_{d-k}\times $S$^{k}$ background, etc.). Each one of
these wave equations is derivable from an effective field theory action in $%
d $ dimensions. These field theory actions look different but yet they all
represents the quantum theory of the same $d+2$ system. Since the original
theory had an SO$\left( d,2\right) $ global symmetry, the derived field
theories, although they look different, must all have SO$\left( d,2\right) $
global symmetry and they must all be related. A well known case of the
symmetry is the conformal SO$\left( d,2\right) $ symmetry of the massless
Klein-Gordon theory. The symmetry must be present for all the others, and
indeed it is the case, provided one takes care of anomalies produced by
quantum ordering of operators. For example, the particle on AdS$_{d-k}\times 
$S$^{k}$ background would not be SO$\left( d,2\right) $ symmetric (for every 
$k$) at the field theory level unless a quantized mass term produced by
quantum ordering is included in the action \cite{lifting}.

Similar comments apply for spin 1/2 wave equations, such as the Dirac
equation, etc. produced by the various gauge fixings of the OSp$\left(
1|2\right) $ gauge theory, or for spin 1 wave equations, such as Maxwell
equation etc. produced by the various gauge fixings of the OSp$\left(
2|2\right) $ theory.

An interesting question is: Is there a master field theory in $d+2$
dimensions from which all of these $d$ dimensional field theories are
derived by a procedure akin to the gauge fixing in the underlying OSp$\left(
n|2\right) $ theory? Furthermore, if field interactions are added to each of
the $d$ dimensional theories, which of these interactions would still
represent the unified master field theory in $d+2$ dimensions, thereby
making the different $d$ dimensional theories all equivalent to each other
under some kind of duality transformation?

These questions are answered by quantizing the worldline theory in a
manifestly SO$\left( d,2\right) $ covariant formalism. The wave equation is
then in $d+2$ dimensions, and it is supplemented by additional field
equations that we call ``kinematic'' as opposed to ``dynamic'' field
equations. The ``kinematic'' equations impose a subset of the underlying OSp$%
\left( n|2\right) $ constraints. The ``dynamic'' field equations correspond
to another subset of constraints, but are derived from a field theory action
in $d+2$ dimensions. Field interactions are included in this dynamic action.
When the kinematic equations are solved, the field theory is reduced from $%
d+2$ dimensions to $d$ dimensions, but there is a choice of which $d$
dimensions among $d+2$ survive in the remaining field equations. This choice
is equivalent to the gauge fixing that could be done in the worldline
formulation of the theory. Indeed the remaining $d$ dimensional field theory
that comes from the $d+2$ field theory correctly produces the wave equations
derived from the gauge fixed worldline theory, including any anomalies. But
now the consistent interactions are also fixed for the $d$ dimensional
version of the theory, since they all come directly from the field
interactions in $d+2$ dimensions.

The formulation of the $d+2$ field equations, both kinematic and dynamic,
proceeds as follows. A physical state $|\Phi >$ of the worldline theory is
labelled by both OSp$\left( n|2\right) $ and SO$\left( d,2\right) $ (if no
background fields) as in (\ref{labels}). The OSp$\left( n|2\right) $ labels
must correspond to a singlet for a gauge invariant physical state. The OSp$%
\left( n|2\right) $ labels include a set of commuting generators in addition
to the OSp$\left( n|2\right) $ Casimir eigenvalues that are zero. On a
physical state that is OSp$\left( n|2\right) $ singlet the SO$\left(
n\right) $ generators given by $\frac{1}{2i}\psi _{\lbrack a}\cdot \psi
_{b]} $ must all vanish (since the physical state must be an SO$\left(
n\right) $ singlet). There is an exception for $n=2$ : the SO$\left(
2\right) $ generator $\frac{1}{2i}\psi _{\lbrack 1}\cdot \psi _{2]}=q$ need
not vanish since every representation of SO$\left( 2\right) $ is a singlet
(although not neutral if $q\neq 0$). In addition, among the set of commuting
operators in OSp$\left( n|2\right) $ that would vanish on a singlet, one is
tempted to choose the generators $P^{2}$ and $P\cdot \psi _{a}$ since these
would produce Klein-Gordon and Dirac equations. If these operators vanish we
would be forced into a free field theory. However, before we impose this
last condition, let us re-examine the expression of the Casimir operator (%
\ref{c2ospn2}) to find out if we can make a weaker choice. As we will see,
this is indeed the case, and the weaker choice will allow us to include
interactions in field theory.

The OSp$\left( n|2\right) $ Casimir (\ref{c2ospn2}) may be rewritten by
pulling $P^{2}$ and $P\cdot \psi _{a}$ to the right side 
\begin{eqnarray}
C_{2}\left( OSp\left( n|2\right) \right) &=&\frac{1}{4}\left( iX\cdot P+%
\frac{d+2}{2}+\left| q\right| \delta _{n,2}\right) \left( iX\cdot P+\frac{d-2%
}{2}+n-\left| q\right| \delta _{n,2}\right)  \label{spn2rewrite} \\
&&+\frac{1}{4}X^{2}P^{2}-\frac{i}{2}X\cdot \psi _{a}P\cdot \psi _{a}+\frac{1%
}{32}\left( \psi _{\lbrack a}\cdot \psi _{b]}\right) \left( \psi _{\lbrack
a}\cdot \psi _{b]}\right)  \nonumber
\end{eqnarray}
To define a physical state, with a vanishing $C_{2}\left( OSp\left(
n|2\right) \right) =0,$ it is sufficient to simultaneously diagonalize the
commuting operators $iX\cdot P,$ $X^{2}P^{2},$ $X\cdot \psi _{a}P\cdot \psi
_{a}$ all of which commute also with the SO$\left( n\right) $ generators $%
\frac{1}{2i}\psi _{\lbrack a}\cdot \psi _{b]}$. Thus, a physical state is
defined by 
\begin{eqnarray}
X^{2}P^{2}|\Phi &>&=0,\quad \left( iX\cdot P+\frac{d-2}{2}+n-\left| q\right|
\delta _{n,2}\right) |\Phi >=0  \label{ppandp} \\
X\cdot \psi _{a}P\cdot \psi _{a}|\Phi &>&=0,\quad \left( \frac{1}{2i}\psi
_{\lbrack a}\cdot \psi _{b]}-q\,\delta _{n,2}\,\,\varepsilon _{ab}\right)
|\Phi >=0\,.  \label{psipsi}
\end{eqnarray}
Demanding an OSp$\left( n|2\right) $ singlet also imposes the SO$\left(
d,2\right) $ Casimir eigenvalue given in (\ref{Cospn2}). Some additional
operators, even if they do not commute with the above, may have definite
eigenvalues on physical states $|\Phi >$, since we are interested in the
states that give only the zero eigenvalues of the operators above rather
than all of their eigenvalues. It may then be quantum mechanically
compatible if certain additional operators take on specific values as well
on the physical states (for example, even though the SO$\left( n\right) $
generators do not commute with each other they can all vanish simultaneously
on a SO$\left( n\right) $ singlet).

In addition to the physical ket states $|\Phi >$ we also consider the spin
and position space bra states $<$ $X,spin|.$ The probability amplitude $%
<X,spin|\Phi >\equiv \Phi _{spin}\left( X\right) $ defines the physical
fields or wavefunctions that will enter in the $d+2$ dimensional field
theory. The spin labels will be explained below. On the state $<X,spin|$ the
position operators $X^{M}$ are diagonal. An important property of this state
is defined by demanding the $X^{2}$ operator to vanish $<X,spin|X^{2}=0$ as
a constraint imposed on the position Hilbert space. From 
\begin{equation}
0=<X,spin|X^{2}|\Phi >\equiv X^{2}\Phi _{spin}\left( X\right)  \label{kin}
\end{equation}
we learn that $\Phi _{spin}\left( X\right) $ vanishes everywhere, except on
the $d+2$ dimensional lightcone where $X^{2}=0.$ Therefore, to examine the
non-trivial fields we must take $X^{2}=0.$ On position space the momentum
operators act as derivatives $<X,spin|P^{M}=-i\partial _{M}<X,spin|.$ The
quantization procedure we have just adopted (i.e. imposing $X^{2}$ on bra
states) implies that when there are derivatives applied on the fields, such
as $\partial _{M}\Phi _{spin}\left( X\right) ,$ the derivative must be
performed first before imposing the constraint 
\begin{equation}
X^{2}=0.  \label{kin1}
\end{equation}
This describes one of the ``kinematic'' equations that will be needed.
Another kinematic constraint is the second equation in (\ref{ppandp}). On
the fields it takes the form 
\begin{equation}
\left( X\cdot \partial +\frac{d-2}{2}+n-\left| q\right| \delta _{n,2}\right)
\Phi _{spin}\left( X\right) =0.  \label{kin2}
\end{equation}
where $\left| q\right| $ will be related to the spin in the case of $n=2.$
Basically this requires fields of specific scales depending on their spin.
The required scale is in $d+2$ dimensions, not in $d$ dimensions. A third
kinematic equation is the second equation in (\ref{psipsi}), but we will
solve that one completely and the fields $\Phi _{spin}\left( X\right) $ will
be defined after the explicit solution of that equation.

There remains the ``dynamic'' equations, the first equations in (\ref{ppandp}%
,\ref{psipsi}), which yield Klein-Gordon or Dirac type equations for the
fields $\Phi _{spin}\left( X\right) .$ In the next few sections we study the
dynamic equations for each spinning field $\Phi _{spin}\left( X\right) ,$
include field interactions, and build an action from which they can be
derived. The combination of the interacting field theory action and the
kinematic equations (\ref{kin1},\ref{kin2}) define the $d+2$ dimensional
field theory at the classical level.

\section{Scalar field (n=0)}

For $n=0$ ( drop $\psi _{a}^{M}$) the worldline theory based on OSp$\left(
0|2\right) =Sp\left( 2,R\right) $ describes a spinless particle. The dynamic
(\ref{ppandp}) and kinematic equations (\ref{kin1},\ref{kin2}) take the form 
\begin{equation}
X^{2}\partial ^{M}\partial _{M}\,\Phi \left( X\right) =0,\quad X^{M}\partial
_{M}\,\Phi \left( X\right) =-\frac{d-2}{2}\,\Phi \left( X\right) ,\quad
X^{2}\,\Phi \left( X\right) =0,  \label{scalar}
\end{equation}
Consistent interactions have the form 
\begin{equation}
\partial ^{M}\partial _{M}\,\Phi =\lambda \Phi ^{\left( d+2\right) /\left(
d-2\right) }+\cdots .  \label{interact}
\end{equation}
where $\cdots $ stands for interactions with other fields that we will
discuss below. All interactions are constrained by demanding consistency
with the Sp$\left( 2,R\right) $ kinematic constraints in (\ref{scalar}),
which are imposed by applying $X\cdot \partial $ or $X^{2}$ on both sides,
and using (\ref{scalar}). Without the interactions this equation is
consistent with choosing to diagonalize $P^{2}\sim 0$ on the physical state,
which was possible in the first place, but by going through the steps above
we see that $\partial ^{M}\partial _{M}\,\Phi \left( X\right) $ need not
vanish while remaining consistent with the physical state conditions. In
general, if written in radial coordinates, the Laplacian operator $\partial
^{M}\partial _{M}$ in $d+2$ dimensions has terms proportional to $1/X^{2},$
which will tend to blow up as $X^{2}\rightarrow 0$ 
\begin{equation}
\partial ^{M}\partial _{M}=\frac{1}{X^{2}}\left( \left( X\cdot \partial
\right) ^{2}+dX\cdot \partial -\frac{1}{2}L^{MN}L_{MN}\right) 
\end{equation}
but the numerator is zero after using the second equation in (\ref{scalar})
and the physical value of the SO$\left( d,2\right) $ Casimir (\ref{C2sod2})
for $n=0$. Therefore the operator $\partial ^{M}\partial _{M}\,\sim 0/0$ is
finite on a physical state as given in (\ref{interact}). In this way we have
seen that the underlying $Sp\left( 2,R\right) $ gauge symmetry permits only
certain interactions. If $d+2=6$ (i.e. $d=4$) the right hand side of (\ref
{interact}) contains $g\Phi ^{3}.$ The field equation can be derived from
the variation of the Lagrangian 
\begin{equation}
L_{d+2}^{\Phi }=-\frac{1}{2}\Phi \partial ^{M}\partial _{M}\Phi -\lambda 
\frac{\left( d-2\right) }{2d}\Phi ^{2d/\left( d-2\right) },  \label{LF}
\end{equation}
and it must be supplemented by the subsidiary kinematic conditions in (\ref
{scalar}).

Evidently one can write a richer $d+2$ field theory involving several scalar
fields that have interactions with each other so long as those interactions
are consistent with the subsidiary kinematic conditions. This means that the
power $2d/\left( d-2\right) $ should be saturated, but this can be done by
the product of several scalar fields. If $d=4$ the interaction if $\Phi ^{4},
$ but other powers are not permitted.

The equations in (\ref{scalar}) are a slight generalization of Dirac's
equations \cite{Dirac} that he obtained by a different set of arguments
(instead of the first eq. in (\ref{scalar}) he had $\partial ^{M}\partial
_{M}\,\Phi \left( X\right) =0$). In our case these equations follow directly
from the Sp$\left( 2,R\right) $ gauge symmetry conditions of the worldline
theory, and thus provide a gauge theory basis for Dirac's approach.

We will next solve the subsidiary kinematic field equations and show that
the remaining dynamics is described by a field theory in $d$ dimensions.
However, we will see that there are many ways of choosing coordinates in
coming from $d+2$ dimensions down to $d$ dimensions while solving the
subsidiary conditions. The choice of coordinates is parallel to fixing a Sp$%
\left( 2,R\right) $ gauge in the worldline theory. Various one-time field
theories in $d$ dimensions emerge when ``time'' is identified in different
ways within the $d+2$ dimensional space. One of those cases corresponds to
conformal field theory, with SO$\left( d,2\right) $ interpreted as the
conformal group, as Dirac suggested. However, all other choices of
coordinates lead to other $d$ dimensional field theories with SO$\left(
d,2\right) $ symmetry, but with SO$\left( d,2\right) $ taking on different
meanings as less familiar hidden symmetries. Thus, the content of these
field equations goes well beyond the linearization of conformal symmetry
envisaged by Dirac and the literature that followed his path \cite{Dirac} 
\cite{Kastrup}\cite{salam}\cite{vasilev}. In fact, the equations above unify
a class of different looking $d$-dimensional one-time field theories into
the same $d+2$ dimensional two-time field theory, including interactions, as
shown below.

\subsection{Massless scalar field in d dimensions}

In the worldline formulation the gauge fixing $X^{+}\left( \tau \right) =1$
and $P^{+}\left( \tau \right) =0,$ and solution of constraints $X^{2}=0$ and 
$X\cdot P=0$ left behind the Minkowski coordinates and momenta $x^{\mu
}\left( \tau \right) ,$ $p^{\mu }\left( \tau \right) $ as the independent
degrees of freedom 
\begin{eqnarray}
X^{+^{\prime }}\left( \tau \right) &=&1,\quad X^{-^{\prime }}\left( \tau
\right) =x^{2}/2,\quad X^{\mu }=x^{\mu }\left( \tau \right) ,  \label{confX}
\\
\quad P^{+^{\prime }}\left( \tau \right) &=&0,\quad P^{-^{\prime }}\left(
\tau \right) =x\cdot p,\quad P^{\mu }=p^{\mu }\left( \tau \right)
\label{confP}
\end{eqnarray}
constrained only by $p^{2}=0.$ The dynamics of the remaining coordinates
describe the massless relativistic particle \cite{conf}. The quantization of
the remaining system produced the Klein-Gordon equation which in turn can be
derived from the Klein-Gordon action that has the SO$\left( d,2\right) $
conformal symmetry identified with the Lorentz symmetry $%
L^{MN}=X^{M}P^{N}-X^{N}P^{M}$ in $d+2$ dimensions \cite{conf}. Field
interactions may then be added, but there is no specific instructions for
which interactions are permitted, unless one tries to maintain the SO$\left(
d,2\right) $ Lorentz symmetry.

Now, let us do the analog of this gauge fixing directly in the $d+2$
dimensional field theory of the previous section. Following Dirac, we use
the change of variables 
\begin{equation}
X^{+^{\prime }}=\kappa ,\quad X^{-^{\prime }}=\kappa \lambda ,\quad X^{\mu
}=\kappa x^{\mu },  \label{confpart}
\end{equation}
where the one-time is embedded in Minkowski space $x^{\mu }$ while the
dependence on the other time will be determined by solving the kinematic
field equations. Using the chain rule, $\partial _{M}=\frac{\partial \kappa 
}{\partial X^{M}}\frac{\partial }{\partial \kappa }+\frac{\partial \lambda }{%
\partial X^{M}}\frac{\partial }{\partial \lambda }+\frac{\partial x^{\mu }}{%
\partial X^{M}}\frac{\partial }{\partial x^{\mu }}$ we find 
\begin{eqnarray}
\frac{\partial }{\partial X^{+^{\prime }}} &=&\frac{\partial }{\partial
\kappa }-\frac{\lambda }{\kappa }\frac{\partial }{\partial \lambda }-\frac{%
x^{\mu }}{\kappa }\frac{\partial }{\partial x^{\mu }}  \label{chain1} \\
\frac{\partial }{\partial X^{-^{\prime }}} &=&\frac{1}{\kappa }\frac{%
\partial }{\partial \lambda },\quad \frac{\partial }{\partial X^{\mu }}=%
\frac{1}{\kappa }\frac{\partial }{\partial x^{\mu }},  \label{chain2}
\end{eqnarray}
Note that $P^{+^{\prime }}$ (which was set to zero as a gauge choice in the
worldline approach) is represented by the derivative operator 
\begin{equation}
P^{+^{\prime }}=-P_{-^{\prime }}=i\frac{\partial }{\partial X^{-^{\prime }}}=%
\frac{1}{\kappa }\frac{\partial }{\partial \lambda }.  \label{pplus}
\end{equation}
At this stage no gauge choices have been made;= only a change to more
convenient coordinates has been performed, but note the parallel with the
gauge in (\ref{confX},\ref{confP}). Next, we can write the differential
operators in the new coordinates 
\begin{eqnarray}
X^{M}\partial _{M} &=&\kappa \frac{\partial }{\partial \kappa },\quad
\label{dim} \\
\partial ^{M}\partial _{M} &=&\frac{1}{\kappa ^{2}}\left( \frac{\partial }{%
\partial x^{\mu }}+x_{\mu }\frac{\partial }{\partial \lambda }\right) ^{2}-%
\frac{1}{\kappa ^{2}}\left( 2\kappa \frac{\partial }{\partial \kappa }%
+d-2\right) \frac{\partial }{\partial \lambda }  \label{laplace} \\
&&+\frac{1}{\kappa ^{2}}\left( 2\lambda -x^{2}\right) \left( \frac{\partial 
}{\partial \lambda }\right) ^{2}  \nonumber
\end{eqnarray}
These differential operators are to be applied on a physical field which is
parametrized as $\Phi \left( \kappa ,\lambda ,x^{\mu }\right) $ {\it before}
imposing the kinematic constraints $X^{2}=0.$ To impose this constraint one
must set $\lambda =x^{2}/2$ after differentiation $\frac{\partial }{\partial
\lambda }$. Then we see that the third term in $\partial ^{M}\partial _{M}$
drops out on sufficiently non-singular wavefunctions. Using the kinematic
constraint in (\ref{scalar}) together with (\ref{dim}) the kappa dependence
is fully determined as an overall factor $\kappa ^{-\left( d-2\right) /2}$%
\begin{equation}
\Phi \left( X\right) =\kappa ^{-\left( d-2\right) /2}f\left( x,\lambda
\right) .
\end{equation}
This solution allows us to drop also the second term in $\partial
^{M}\partial _{M}$. Next, note that derivatives with respect to $x^{\mu }$
appear only in the combination $\frac{\partial }{\partial x^{\mu }}+x_{\mu }%
\frac{\partial }{\partial \lambda }.$ Then, setting $\lambda =x^{2}/2$ after
differentiation using the derivative operator $\frac{\partial }{\partial
x^{\mu }}+x_{\mu }\frac{\partial }{\partial \lambda }$ gives the same result
as setting $\lambda =x^{2}/2$ before differentiation and differentiating
only with $\frac{\partial }{\partial x^{\mu }}$%
\begin{equation}
\left[ \left( \frac{\partial }{\partial x^{\mu }}+x_{\mu }\frac{\partial }{%
\partial \lambda }\right) f\left( x,\lambda \right) \right] _{\lambda
=x^{2}/2}=\frac{\partial }{\partial x^{\mu }}f\left( x,\frac{x^{2}}{2}%
\right) .
\end{equation}
Therefore we can set $f\left( x,\lambda \right) |_{\lambda =x^{2}/2}=\phi
\left( x\right) $ before differentiation provided we also drop the term $%
\frac{\partial }{\partial \lambda }$ in the derivative operator $\frac{%
\partial }{\partial x^{\mu }}+x_{\mu }\frac{\partial }{\partial \lambda }.$
We see that all $\frac{\partial }{\partial \lambda }$ terms have dropped out
from the Laplace operator $\partial ^{M}\partial _{M}$ in (\ref{laplace}).
The disappearance of $\frac{\partial }{\partial \lambda }$ everywhere is
parallel to setting $P^{+^{\prime }}=0$ as a Sp$\left( 2,R\right) $ gauge
choice as in (\ref{confP},\ref{pplus}). Using these remarks we see that the
physical state conditions (\ref{scalar},\ref{interact}) are by now fully
solved in this gauge by the following general form 
\begin{equation}
\Phi \left( X\right) =\kappa ^{-\left( d-2\right) /2}\phi \left( x\right)
,\quad \frac{\partial ^{2}\phi \left( x\right) }{\partial x^{\mu }\partial
x_{\mu }}=\lambda \phi ^{\left( d+2\right) /\left( d-2\right) },  \label{kg}
\end{equation}
where $\phi \left( x\right) $ is an interacting massless Klein-Gordon field
in $d$-dimensional Minkowski spacetime (in this interaction we assumed a
single real field, but it could be more general). The effective action that
generates this equation of motion is 
\begin{equation}
L_{d}^{\phi }=\,-\frac{1}{2}\phi \partial ^{\mu }\partial _{\mu }\phi -\frac{%
\lambda \left( d-2\right) }{8d}\phi ^{2d/\left( d-2\right) }.  \label{kgact}
\end{equation}
This is in full agreement with the effective field theory that was obtained
by quantizing the worldline formalism in the fixed gauge $X^{+^{\prime
}}\left( \tau \right) =1,$ $P^{+^{\prime }}\left( \tau \right) =0,$ as given
in \cite{conf}.

Note that the $d+2$ Lagrangian (\ref{LF}) reduces directly to the $d$
Lagrangian (\ref{kgact}) when the solution of the subsidiary conditions (\ref
{kg}) and the form of the Laplacian (\ref{laplace}) are used 
\begin{equation}
L_{d+2}^{\Phi }\left( X\right) \rightarrow \kappa ^{-d}L_{d}^{\phi }\left(
x\right)  \label{kgL}
\end{equation}
$\kappa $ disappears after integration over $\kappa $ in the action.

Thus, solving just the kinematic equations $X^{2}=0$ and $X\cdot \partial
\Phi =-\frac{1}{2}\left( d-2\right) \Phi $ with a particular choice of the
remaining $d$ coordinates, and replacing the solution into the SO$\left(
d,2\right) $ invariant action is sufficient to obtain the dynamics and the
interpretation of the theory in $d$ dimensions.

It is well known that the interacting massless Klein-Gordon theory (\ref
{kgact}), including the interaction, is invariant under conformal
transformations, although the symmetry is somewhat ``hidden''. In the two
time formalism given above the conformal symmetry is inherited from the
manifestly SO$\left( d,2\right) $ invariant equations (\ref{scalar},\ref
{interact}) as shown in different ways in \cite{Dirac} and \cite{conf}. This
allows us to interpret conformal symmetry in $d$ dimensions as the {\it %
Lorentz symmetry} in $d+2$ dimensions acting on the space $X^{M}.$

Thus conformal symmetry in (\ref{kg}) can be taken as evidence for an
underlying higher space with one more timelike and one more spacelike
dimensions. In this higher spacetime all $d+2$ dimensions are at an equal
footing - it is only because of the asymmetric choice of coordinates $\kappa
,\lambda ,x^{\mu }$ that (i) the remaining one ``time'' $x^{0}$ was defined
and (ii) the manifest symmetry was broken artificially in the process of
solving the ``kinematic'' equations (gauge constraints) to rewrite the $d+2$
field equations as a field theory in $d$ dimensions.

The more unifying aspect of the higher space, and the interpretation of the
hidden symmetry as being simply the higher Lorentz symmetry, will make a
stronger impression on the reader after noting that a similar observation is
repeated in several seemingly unrelated field theoretic models that are
actually derivable from the same set of field theoretic equations in the
higher dimensions. Each of the derived field theories in $d$ dimensions has
the SO$\left( d,2\right) $ symmetry realized in the {\it same irreducible
unitary representation}, but its interpretation is not conformal symmetry.
Nevertheless, it is the same Lorentz symmetry of the higher $d+2$ spacetime.

\subsection{Scalar field in AdS$_{D}\times S^{k}$ background}

To show that the meaning of SO$\left( d,2\right) $ goes beyond the conformal
symmetry interpretation, let us now demonstrate that the same SO$\left(
d,2\right) $ invariant equations (\ref{scalar},\ref{interact}) have a
different physical interpretation when the coordinates, in particular
``time'', are chosen in a different way. Let the $d+2=D+k+2$ coordinates be
labelled as $X^{M}=\left( X^{+^{\prime }},X^{-^{\prime }},X^{\mu
},X^{i}\right) $ with $X^{\mu }$ representing $\left( D-1\right) $ spacetime
dimensions with one time, and $X^{i}$ representing $k+1$ spacelike
dimensions, so $D+k=d$. Consider the following change of variables (this is
related to the Sp$\left( 2,R\right) $ gauge choice for a particle moving in
the AdS$_{D}\times S^{k}$ background in the worldline formalism as given in 
\cite{lifting}) 
\begin{eqnarray}
X^{+^{\prime }} &=&\rho u,\quad X^{-^{\prime }}=\rho \sigma ,\quad
X^{i}=\rho \frac{{\bf u}^{i}}{u}a,\quad X^{\mu }=\frac{1}{a}\rho ux^{\mu }.
\label{adspart} \\
\rho &=&\frac{\sqrt{X_{i}^{2}}}{a},\quad \sigma =\frac{aX^{-^{\prime }}}{%
\sqrt{X_{i}^{2}}},\quad {\bf u}^{i}=\frac{aX^{+^{\prime }}X^{i}}{X_{i}^{2}}%
,\quad x^{\mu }=\frac{aX^{\mu }}{X^{+^{\prime }}}
\end{eqnarray}
The ${\bf u}^{i}$ are Euclidean vectors in $k+1$ dimensions, $u$ is the
magnitude of the Euclidean vector $u=\left| {\bf u}\right| ,$ $a$ is a
constant with dimension of length, and $x^{\mu }$ are Minkowski vectors in $%
\left( D-1\right) $ dimensions. The $X^{2}=0$ condition gives 
\begin{equation}
\sigma =\frac{a^{4}+x^{2}u^{2}}{2ua^{2}}.  \label{lads}
\end{equation}
The SO$\left( d,2\right) $ covariant line element in $d+2$ dimensions $%
dX\cdot dX$ gives the AdS$_{D}\times S^{k}$ line element in $D+k=d$
dimensions up to a conformal factor (after using (\ref{lads})), 
\begin{eqnarray}
dX\cdot dX &=&\rho ^{2}\left( \frac{\left( d{\bf u}\right) ^{2}}{u^{2}}+%
\frac{u^{2}}{a^{2}}\left( dx_{\mu }\right) ^{2}\right)  \label{adsmetric} \\
&=&\rho ^{2}\left( \left( d\Omega _{k}\right) ^{2}+\frac{du^{2}}{u^{2}}+%
\frac{u^{2}}{a^{2}}\left( dx_{\mu }\right) ^{2}\right) ,
\end{eqnarray}
where $\left( d\Omega _{k}\right) ^{2}$ is the metric on $S^{k}.$ This shows
the relationship of the parametrization to the AdS$_{D}\times S^{k}$
background, with $D+k=d$. We will consider all possible values of $%
k=0,1,\cdots ,\left( d-2\right) ,$ so that we will exhibit a relation among
the field theories for fixed $d$, written on the backgrounds AdS$_{d},$ AdS$%
_{d-1}\times S^{1},\cdots ,$ AdS$_{2}\times S^{d-2}$.

Let us rewrite (\ref{scalar},\ref{interact}) in these coordinates. The chain
rule $\partial _{M}=\left( \partial _{M}\rho \right) \frac{\partial }{%
\partial \rho }+\left( \partial _{M}\sigma \right) \frac{\partial }{\partial
\sigma }+\left( \partial _{M}{\bf u}^{i}\right) \frac{\partial }{\partial 
{\bf u}^{i}}+\left( \partial _{M}x^{\mu }\right) \frac{\partial }{\partial
x^{\mu }}$ gives 
\begin{eqnarray}
\frac{\partial }{\partial X^{+^{\prime }}} &=&\frac{1}{\rho u}\left( {\bf u}%
^{i}\frac{\partial }{\partial {\bf u}^{i}}-x^{\mu }\frac{\partial }{\partial
x^{\mu }}\right) ,\quad \frac{\partial }{\partial X^{-^{\prime }}}=\frac{1}{%
\rho }\frac{\partial }{\partial \sigma },\quad \frac{\partial }{\partial
X^{\mu }}=\frac{a}{\rho u}\frac{\partial }{\partial x^{\mu }} \\
\frac{\partial }{\partial X^{i}} &=&\frac{{\bf u}^{i}}{au}\left( \frac{%
\partial }{\partial \rho }-\frac{\sigma }{\rho }\frac{\partial }{\partial
\sigma }-2\frac{{\bf u}^{j}}{\rho }\frac{\partial }{\partial {\bf u}^{j}}%
\right) +\frac{u}{a\rho }\frac{\partial }{\partial {\bf u}^{i}}
\end{eqnarray}
Using these, the relevant differential operators $X^{M}\partial _{M},$ $%
\partial _{M}\partial ^{M}$ (before using (\ref{lads})) may be written in
the form 
\begin{eqnarray}
X^{M}\partial _{M} &=&\rho \frac{\partial }{\partial \rho }  \label{xdads} \\
\partial _{M}\partial ^{M} &=&\frac{a^{2}}{\rho ^{2}u^{2}}\left( D_{\mu
}\right) ^{2}+\left[ \frac{{\bf u}^{i}}{au}\frac{\partial }{\partial \rho }+%
\frac{u}{a\rho }\left( {\bf D}_{i}-2\frac{{\bf u}^{i}}{u}\frac{{\bf u}^{j}}{u%
}{\bf D}_{j}\right) \right] ^{2}+\cdots  \label{ddads}
\end{eqnarray}
where the derivative operators $D_{\mu },{\bf D}_{i}$ are given by 
\begin{equation}
D_{\mu }=\frac{\partial }{\partial x^{\mu }}+\frac{ux_{\mu }}{a^{2}}\frac{%
\partial }{\partial \sigma },\quad {\bf D}_{i}=\frac{\partial }{\partial 
{\bf u}^{i}}+\frac{{\bf u}^{i}}{u}\left( \frac{x^{2}}{2a^{2}}-\frac{a^{2}}{%
2u^{2}}\right) \frac{\partial }{\partial \sigma },
\end{equation}
and the terms $\cdots $ are all proportional to $\left( 2ua^{2}\sigma
-a^{4}-x^{2}u^{2}\right) $ which vanishes according to (\ref{lads}).

The general solution of the second equation in (\ref{scalar}) now takes the
form 
\begin{equation}
\Phi \left( X\right) =\rho ^{-\left( d-2\right) /2}F\left( \sigma ,{\bf u,}%
x\right) |_{\sigma =\left( a^{4}+x^{2}u^{2}\right) /2a^{2}u}.
\end{equation}
We note again that it is possible to replace the differential operators $%
D_{\mu },{\bf D}_{i}$ that are applied before the substitution $\sigma
=\left( a^{4}+x^{2}u^{2}\right) /2ua^{2}$ with the simple differentiation $%
\frac{\partial }{\partial x^{\mu }},\frac{\partial }{\partial {\bf u}^{i}}$
if the substitution is done before differentiation 
\begin{eqnarray}
\left[ D_{\mu }F\left( \sigma ,{\bf u,}x\right) \right] _{\sigma =\left(
a^{4}+x^{2}u^{2}\right) /2ua^{2}} &=&\frac{\partial }{\partial x^{\mu }}%
F\left( \frac{a^{4}+x^{2}u^{2}}{2ua^{2}},{\bf u,}x\right) \\
\left[ {\bf D}_{i}F\left( \sigma ,{\bf u,}x\right) \right] _{\sigma =\left(
a^{4}+x^{2}u^{2}\right) /2ua^{2}} &=&\frac{\partial }{\partial {\bf u}^{i}}%
F\left( \frac{a^{4}+x^{2}u^{2}}{2ua^{2}},{\bf u,}x\right)
\end{eqnarray}
Therefore, we may define the field $\phi \left( x,{\bf u}\right) $ that
depends only on the AdS$_{d-k}\times S^{k}$ variables $x^{m}=\left( x^{\mu },%
{\bf u}^{i}\right) $%
\begin{equation}
\phi \left( x,{\bf u}\right) \equiv F\left( \sigma ,{\bf u,}x\right)
|_{\sigma =\left( a^{4}+x^{2}u^{2}\right) /2a^{2}u}
\end{equation}
Combined with the vanishing of the $\cdots $ terms in (\ref{ddads}) the net
effect is to drop the derivatives $\partial /\partial \sigma $ wherever they
appear. This is equivalent to the choice of the Sp$\left( 2,R\right) $ gauge 
$P^{+^{\prime }}=\frac{\partial }{i\partial X^{-^{\prime }}}=\frac{1}{i\rho }%
\frac{\partial }{\partial \sigma }=0$ which was performed in the worldline
formalism \cite{lifting}. With these remarks, we then find that the full set
of equations (\ref{scalar}, \ref{interact}, \ref{xdads}, \ref{ddads}) are
solved provided $\phi \left( x,{\bf u}\right) $ satisfies the scalar
equation in the AdS$_{d-k}\times S^{k}$ background with a quantized mass
term 
\begin{eqnarray}
\Phi \left( X\right) &=&\rho ^{-\left( d-2\right) /2}\phi \left( x,{\bf u}%
\right)  \label{adsf} \\
0 &=&\frac{1}{\sqrt{-G}}\partial _{m}\left( \sqrt{-G}G^{mn}\partial _{n}\phi
\right) +M^{2}\phi +\lambda \phi ^{\left( d+2\right) /\left( d-2\right) }
\label{adsseq} \\
M^{2} &\equiv &\frac{1}{4a^{2}}\left( d-2\right) \left( d-2k\right) ,
\label{mass}
\end{eqnarray}
where the metric $G_{mn}$ is given by the AdS$_{d-k}\times S^{k}$ line
element, with labels $x^{m}=\left( x^{\mu },{\bf u}^{i}\right) $ 
\begin{equation}
ds^{2}=\frac{\left( d{\bf u}\right) ^{2}}{u^{2}}+\frac{u^{2}}{a^{2}}\left(
dx_{\mu }\right) ^{2}\equiv G_{mn}dx^{m}dx^{n}.
\end{equation}
Note that the mass term vanishes if $d=2$ or if $d=2k.$ Thus for AdS$%
_{2}\times S^{d-2}$ and AdS$_{d/2}\times S^{d/2}$ there is no mass term.
These equations follow from the Lagrangian in $d$ total dimensions 
\begin{eqnarray}
L_{d}^{\phi } &=&{\bf \,}-\frac{1}{2}\phi \partial _{m}\left( \sqrt{-G}%
G^{mn}\,\partial _{n}\phi \right)  \label{adsslag} \\
&&-\sqrt{-G}\left[ \frac{\left( d-2\right) \left( d-2k\right) }{8a^{2}}\phi
^{2}+\frac{\lambda \left( d-2\right) }{2d}\phi ^{2d/\left( d-2\right) }%
\right]
\end{eqnarray}
This Lagrangian also follows directly from the $d+2$ dimensional Lagrangian
by inserting the solution of the kinematic constraints given in (\ref{adsf}) 
\begin{equation}
L_{d+2}^{\Phi }=\rho ^{-d}L_{d}^{\phi }.
\end{equation}
The $\rho $ dependence disappears in the action after an integration of the
Lagrangian in $d+2$ dimensions.

The same result was derived in \cite{lifting} by choosing a Sp$\left(
2,R\right) $ gauge in the worldline formalism and then doing non-covariant
quantization. There, it was essential to figure out the correct ordering of
the quantum operators, which in turn gave rise to the quantized mass given
above. Thus, the quantized mass term is a {\it quantum anomaly}. If the
anomaly is missed, the AdS$_{d-k}\times $S$^{k}$ theory would no longer be
equivalent to the $d+2$ dimensional theory or any of the other $d$
dimensional versions.

The evident symmetry of this action is only SO$\left( d-k-1,2\right) \times
SO\left( k+1\right) $ which corresponds to the Killing symmetries of the AdS$%
_{d-k}\times S^{k}$ metric $G_{mn}.$ However, there is more hidden symmetry
in this action that was not noticed before the advent of two-time physics 
\cite{lifting}. In the present field theory setting this follows simply from
the property that the original set of equations (\ref{scalar}, \ref{interact}%
) are invariant under the larger SO$\left( d,2\right) .$ This contains the
Killing symmetries as a subgroup, but the total symmetry is larger.
Therefore we should expect that there are hidden symmetries in the effective
action that correspond to the additional generators in the coset 
\begin{equation}
\frac{SO\left( d,2\right) }{SO\left( d-k-1,2\right) \times SO\left(
k+1\right) }.
\end{equation}
That is, the effective action given above should have the full SO$\left(
d,2\right) =SO\left( D+k,2\right) $ symmetry for every $k$. Indeed it was
shown in \cite{lifting}, that this action has the full symmetry SO$\left(
d,2\right) .$ The quantized mass term is essential for this symmetry to be
valid. Hence, the larger symmetry requires a quantized mass. The generators
of the full symmetry, and the transformation of $\phi \left( x,{\bf u}%
\right) $ under them are explicitly given for every $k$ in \cite{lifting}.
The presence of the symmetry is again evidence for the underlying larger
space that contains one more spacelike and more timelike dimensions.

Through this example, with various $k$, we have demonstrated that the
content of the fully covariant equations (\ref{scalar}, \ref{interact}) is
much more than the conformal massless particle that was originally aimed for
by Dirac \cite{Dirac}. The field theoretic results reported here fully agree
with the worldline formalism at the quantum level performed also at fixed
gauges\cite{lifting}. Furthermore, in the field theory formalism field
interactions consistent with the SO$\left( d,2\right) $ symmetry are also
introduced directly in $d+2$ theory.

It is interesting to consider the AdS-CFT correspondence \cite{maldacena}- 
\cite{witten} in this setting. Going to the boundary of AdS corresponds to $%
u\rightarrow \infty .$ In this limit the original form of the theory in $d+2$
dimensions can be analyzed easily by examining the parametrization given in (%
\ref{adspart}). We may also define $\rho =\kappa /u$ to more easily extract
the information when we take the limit with finite $\kappa $. In this limit
the coordinates and momenta have the form 
\begin{equation}
X^{+^{\prime }}\rightarrow \kappa ,\quad X^{-^{\prime }}\rightarrow \frac{%
\kappa x^{2}}{2a^{2}},\quad X^{i}\rightarrow 0,\quad X^{\mu }\rightarrow 
\frac{\kappa }{a}x^{\mu }.
\end{equation}
We see that the $d+2$ space shrinks in the $k+1$ dimensions $X^{i},$ and
remains finite in the $d-k-1$ dimensions $X^{\mu }.$ Then the two-time
Lagrangian (\ref{LF}) gets reduced $L_{d+2}^{\Phi }\rightarrow
L_{d-k+1}^{\Phi }$ in the number of dimensions. By comparison to the
parametrization of the previous section, and recalling eqs.(\ref{kg}-\ref
{kgL}), we learn that the full field theory given by $L_{d+2}^{\Phi }$ now
shrinks to a conformal field theory in $d-k-1$ dimensions that defines the
boundary of the AdS space 
\begin{equation}
L_{d+2}^{\Phi }\rightarrow L_{d-k+1}^{\Phi }=\kappa ^{\left( d-k-1\right)
/2}L_{d-k-1}^{\phi }.
\end{equation}
This is precisely the AdS-CFT correspondence applied to this theory. Having
the two-time theory in the form $L_{d+2}^{\Phi }$ as the common link for
various parametrizations, permitted the analysis to proceed in a
straightforward manner in proving the AdS-CFT correspondence in the present
case.

\subsection{Non-relativistic Schr\"{o}dinger field}

The two cases, massless Klein-Gordon and particle in AdS$\times $S discussed
in the two previous sections are relatively easy from the point of view of
operator ordering in the first quantized theory. In this section we would
like to discuss a harder case in which it is not a priori evident how to
order quantum operators.

In the worldline theory the gauge fixing $P^{+^{\prime }}\left( \tau \right)
=m,$ and $P^{0}\left( \tau \right) =0$ at the classical level produces the
non-relativistic massive particle with mass $m$ \cite{lifting}. In this
gauge the remaining degrees of freedom are designated by the canonical pairs 
$\left( t\left( \tau \right) ,H\left( \tau \right) \right) $ and $\left( 
{\bf r}^{i}\left( \tau \right) ,{\bf p}^{i}\left( \tau \right) \right) $
which are constrained by $H={\bf p}^{2}/2m.$ These are related to the $%
X^{M},P^{M}$ which satisfy $X^{2}=0$ and $X\cdot P=0$ as follows 
\begin{equation}
P^{+^{\prime }}=m,\quad P^{-^{\prime }}=H\left( \tau \right) ,\quad
P^{0}=0,\quad P^{i}={\bf p}^{i}\left( \tau \right)  \label{nonrelP}
\end{equation}
where $m$ is a $\tau $ independent constant, and 
\begin{eqnarray}
X^{+^{\prime }} &=&t\left( \tau \right) ,\quad X^{-^{\prime }}=\frac{1}{m}%
\left( {\bf r\cdot p-}tH\right) ,\quad X^{i}={\bf r}^{i}\left( \tau \right) ,
\label{nonrelX} \\
X^{0} &=&\pm \sqrt{{\bf r}^{2}-\frac{2t}{m}\left( {\bf r\cdot p-}tH\right) }
\label{ex0}
\end{eqnarray}
The $H={\bf p}^{2}/2m$ condition follows from the remaining dynamical
constraint $P^{2}=0.$ Evidently the field theory version of this dynamical
constraint is the Schr\"{o}dinger equation 
\begin{equation}
i\frac{\partial \phi \left( t,{\bf r}\right) }{\partial t}=-\frac{1}{2m}{\bf %
\nabla }^{2}\phi \left( t,{\bf r}\right) +\cdots  \label{scheq}
\end{equation}
that follows from the free Lagrangian in $d$ dimensions 
\begin{equation}
L_{d}^{\phi }=i\phi ^{\ast }\frac{\partial \phi }{\partial t}-\frac{1}{2m}%
{\bf \nabla }\phi ^{\ast }{\bf \nabla }\phi +\cdots
\end{equation}
The dots $\cdots $ represent interactions that could be added.

The non-relativistic particle action $S=\int dt$ $\frac{m}{2}\left( \partial
_{t}{\bf r}\right) ^{2}$ has a surprising SO$\left( d,2\right) $ symmetry
(non-conformal) given by the gauge fixed form of the global SO$\left(
d,2\right) $ Lorentz generators $L^{MN}=X^{M}P^{N}-X^{N}P^{M}$ as explained
in \cite{lifting}. Evidently the field theory that is derived from the $d+2$
field theory must also inherit this symmetry. Operator ordering of the
quantity $X^{0}$ (\ref{ex0}) is non-trivial, and therefore constructing the
SO$\left( d,2\right) $ symmetry generators $L^{MN}$ at the quantum level in
this fixed gauge is not easy. These $L^{MN}$ would be the Noether charges
for the symmetry SO$\left( d,2\right) $ of the Schr\"{o}dinger theory. The
corresponding problem in the previous two cases were solved satisfactorily
by fixing the correct anomalies \cite{lifting}, but the non-linear form of (%
\ref{ex0}) has discouraged the analysis so far. How does this problem show
up in the field theory version, and how is it resolved, in particular when
there are field interactions? Without a guiding symmetry such as SO$\left(
d,2\right) $ there would not be restrictions on the interactions.

Let us now try to imitate directly in the $d+2$ dimensional field theory the
gauge fixing $P^{+^{\prime }}\left( \tau \right) =m,$ and $P^{0}\left( \tau
\right) =0$ of the worldline theory. Before applying any of the kinematic
constraints, the relevant SO$\left( d,2\right) $ differential operators can
be rewritten in the form\footnote{%
Bo Zhang first constructed the following formulas. I thank him for showing
me his work.} 
\begin{equation}
X^{M}\partial _{M}=X^{+^{\prime }}D_{+^{\prime }}+X^{-^{\prime
}}D_{-^{\prime }}+X^{i}D_{i}-X^{M}X_{M}\frac{1}{X^{0}}\partial _{0}
\label{schxd}
\end{equation}
and 
\begin{equation}
\partial ^{M}\partial _{M}=-2D_{+^{\prime }}D_{-^{\prime }}+\left(
D_{i}\right) ^{2}-\frac{2}{X^{0}}\partial _{0}\left( X^{M}\partial _{M}+%
\frac{d-2}{2}\right) -X^{M}X_{M}\left( \frac{1}{X^{0}}\partial _{0}\right)
^{2}  \label{schdd}
\end{equation}
where 
\begin{equation}
D_{+^{\prime }}=\partial _{+^{\prime }}-\frac{X^{-^{\prime }}}{X^{0}}%
\partial _{0},\quad D_{-^{\prime }}=\partial _{-^{\prime }}-\frac{%
X^{+^{\prime }}}{X^{0}}\partial _{0},\quad D_{i}=\partial _{i}+\frac{X_{i}}{%
X^{0}}\partial _{0}.
\end{equation}
Imposing $X^{M}X_{M}=0$ is equivalent to setting 
\begin{equation}
X_{0}=\pm \sqrt{X^{i}X_{i}-2X^{+^{\prime }}X^{-^{\prime }}},  \label{x0}
\end{equation}
but, before doing so, we must apply all the derivatives $\partial _{0}$ on
the wavefunction $\Phi \left( X^{0},X^{+^{\prime }},X^{-^{\prime
}},X^{i}\right) .$ However, from (\ref{schxd},\ref{schdd}) we see that when
the kinematic constraints are applied all terms containing the explicit $%
\partial _{0}$ drop out, except those appearing in the definition of $%
D_{+^{\prime }},D_{-^{\prime }},D_{i}$. Furthermore, for these special
combinations, applying first the derivative and then imposing (\ref{x0})
gives the same result as first imposing (\ref{x0}) and replacing $%
D_{+^{\prime }},D_{-^{\prime }},D_{i}$ with ordinary derivatives $\partial
_{+^{\prime }},\partial _{-^{\prime }},\partial _{i}$%
\begin{equation}
\left[ D_{i,\pm ^{\prime }}\Phi \left( X^{0},X^{+^{\prime }},X^{-^{\prime
}},X^{i}\right) \right] _{X^{0}=\pm \sqrt{X^{i}X_{i}-2X^{+^{\prime
}}X^{-^{\prime }}}}=\partial _{i,\pm ^{\prime }}\Phi |_{X^{0}}
\end{equation}
where we have defined the notation 
\begin{equation}
\Phi |_{X^{0}}\equiv \Phi \left( \pm \sqrt{X^{i}X_{i}-2X^{+^{\prime
}}X^{-^{\prime }}},X^{+^{\prime }},X^{-^{\prime }},X^{i}\right) .
\end{equation}
In this way $\partial _{0}$ completely disappears and $X^{0}$ is expressed
in terms of the other coordinates. This is the field theory version of the
gauge condition $P^{0}\left( \tau \right) =0$ used in the worldline approach.

The next step is to work in a basis that corresponds to $P^{+^{\prime
}}=-P_{-^{\prime }}=m$ while at the same time solve the remaining kinematic
constraint that now takes the form 
\begin{equation}
\left( X^{+^{\prime }}\partial _{+^{\prime }}+X^{-^{\prime }}\partial
_{-^{\prime }}+X^{i}\partial _{i}+\frac{d-2}{2}\right) \Phi |_{X^{0}}=0.
\label{schdim}
\end{equation}
This is done by first going to Fourier space in the $X^{-^{\prime }}$
coordinate and then imposing the kinematic constraint. The result is 
\begin{equation}
\Phi |_{X^{0}}=\int dm\,e^{-imX^{-^{\prime }}}m^{\left( d-4\right) /2}\phi
\left( mX^{+^{\prime }},mX^{i}\right) ,  \label{phitosch}
\end{equation}
where the function $\phi \left( t,{\bf r}^{i}\right) $ is arbitrary. Finally
we apply the dynamical operator on this form and find the Schr\"{o}dinger
operator 
\begin{eqnarray}
\left( \partial ^{M}\partial _{M}\Phi \right) || &=&\left( -2\partial
_{+^{\prime }}\partial _{-^{\prime }}+\partial _{i}\partial ^{i}\right)
\left( \Phi |_{X^{0}}\right) \\
&=&\int dm\,e^{-imX^{-^{\prime }}}m^{\left( d-4\right) /2}\left[ \left( 2im%
\frac{\partial }{\partial t}+{\bf \nabla }_{{\bf r}}^{2}\right) \phi \left(
t,{\bf r}\right) \right] _{t=mX^{+^{\prime }},{\bf r}^{i}{\bf =}mX^{i}}
\end{eqnarray}
On the left side the notation $\left( \partial ^{M}\partial _{M}\Phi \right)
||$ implies that both kinematic constraints have been implemented.

Now we see that the free field equation in $d+2$ dimensions $\left( \partial
^{M}\partial _{M}\Phi \right) ||=0$ corresponds to the free non-relativistic
Schr\"{o}dinger equation with mass $P^{+^{\prime }}=m$%
\begin{equation}
i\frac{\partial }{\partial t}\phi \left( t,{\bf r}\right) =-\frac{1}{2m}{\bf %
\nabla }_{{\bf r}}^{2}\phi \left( t,{\bf r}\right) ,
\end{equation}
in agreement with the first quantization of the worldline theory given in (%
\ref{scheq}). By rewriting it in the form $\left( \partial ^{M}\partial
_{M}\Phi \right) ||=0$ the hidden SO$\left( d,2\right) $ symmetry of the
Schr\"{o}dinger equation is exposed. The interactions consistent with the SO$%
\left( d,2\right) $ symmetry follow from the original equations in $d+2$
dimensions, but unfortunately they do not seem to have a simple or
recognizable form in this case, so we will not discuss it any longer in this
paper.

\subsection{Generalizations}

As argued above, a class of one-time physics dynamics is unified by the
field theoretic two-time formalism. The class is much larger than the cases
discussed above since, as we know from the worldline approach, it includes
other one-time dynamics such as the H-atom, harmonic oscillator, particle in
various potentials, etc.\cite{lifting}. It would be interesting to explore
the interacting field theory for some of these cases. The interaction term
then provides a field theoretic approach to the interaction of these systems
in a setting which has never been explored before. In some generalized sense
this is analogous to duality in M-theory.

The effective one-time field theories thus obtained, the ones in the
previous sections, as well as any others derived similarly in other
embeddings of $d$ dimensions inside the $d+2$ spacetime, are all
representatives of the same two-time field theory which provides for some
remarkable relations among them. Such relations were not apparent before the
insight provided by two-time physics \cite{conf}-\cite{emgrav}. In
principle, in related $d$ dimensional field theories one should be able to
compute Sp$\left( 2,R\right) $ gauge invariant quantities and obtain the
same result. The SO$\left( d,2\right) $ symmetry properties are Sp$\left(
2,R\right) $ gauge invariant, in particular the SO$\left( d,2\right) $ is
realized in the same unitary representation in all the derived $d$%
-dimensional theories. Likewise, it must be possible to compute various Sp$%
\left( 2,R\right) $ gauge invariant quantities and obtain the same or
related results by using the different one-time field theories, including
the interactions. Further computations along these lines, using the full
power of interacting field theory, would help to strenghthen the case for
two time physics, and perhaps help discover some of its utility by
demonstrating that one could perform certain computations more easily by
choosing a particular version of the field theory.

\section{Spin 1/2 field}

If we take $n=1$ in (\ref{ppandp},\ref{psipsi}) then the physical state
describes a spin 1/2 field. The fermion $\psi ^{M}$ is represented by a
Dirac gamma matrix $\psi ^{M}=\gamma ^{M}/\sqrt{2},$ and position space now
has an additional SO$\left( d,2\right) $ spinor index $<X,\alpha |$. 
The fermionic field is given by the probability
amplitude $<X,\alpha |\Phi >=\Psi _{\alpha }\left(
X\right) $. To satisfy the singlet OSp$\left( 1|2\right) $ conditions given
in (\ref{ppandp},\ref{psipsi}) we see that it is sufficient to impose the
kinematic constraints 
\begin{equation}
\quad \left( X\cdot \partial +\frac{d}{2}\right) \Psi _{\alpha }=0,\quad
X^{2}\Psi _{\alpha }=0.  \label{spinhalf}
\end{equation}
and the free field equation $\left( \gamma \cdot X\,\,\gamma \cdot \partial
\Psi \right) _{\alpha }=0.$ The second kinematic constraint follows from the
property of the bra $<X,\alpha |X^{2}=0$. From 
\begin{equation}
\left( \gamma \cdot X\,\gamma \cdot \partial \right) ^{2}=-X^{2}\partial
^{2}+2\left( \gamma \cdot X\right) \,\left( \gamma \cdot \partial \right)
\,\left( X\cdot \partial +\frac{d}{2}\right) ,
\end{equation}
we see $X^{2}\partial ^{2}\Psi _{\alpha }=0$ need not be imposed as a
separate free field equation. To include interactions consistently with the
``kinematic'' constraints we assume that the worldline version of the OSp$%
\left( 1|2\right) $ gauge theory is properly generalized by including
background fields as in \cite{emgrav}. This permits the addition of source
terms to the free field equation 
\begin{equation}
\left[ \gamma \cdot X\,\,\gamma \cdot \left( \partial -iA\right) \Psi \right]
_{\alpha }=h\Phi ^{2/\left( d-2\right) }\left( \gamma \cdot X\,\Psi \right)
_{\alpha }+\left( \gamma \cdot X\,\xi \right) _{\alpha }.  \label{spin1/2}
\end{equation}
where $\xi _{\alpha }$ is any other fermion that does not blow up when $%
X^{2}\rightarrow 0$, and whose dimension is $\left( X\cdot \partial +\frac{%
d-2}{2}\right) \xi _{\alpha }=0.$ The interacting field equation follows
from varying the following Lagrangian 
\begin{equation}
L_{d+2}^{\Psi }=\bar{\Psi}\gamma \cdot X\,\,\gamma \cdot \left( \partial
-iA\right) \Psi -h\Phi ^{2/\left( d-2\right) }\bar{\Psi}\gamma \cdot X\,\Psi
+\bar{\Psi}\gamma \cdot X\,\xi .  \label{Lpsi}
\end{equation}
The inclusion of the Yang-Mills gauge field $A_{M}\left( X\right) $ assumes
that $\Psi $ is charged under the Yang-Mills local internal symmetry. The
scalar $\Phi $ must also have the correct charges to couple to the fermion
with a non-zero coupling constant $h,$ so the notation is schematic. We also
assumed that the field $\Phi $ included on the right hand side may be of the
type described in the previous section; if so this coupling would modify the
field equations for $\Phi $ given in the previous section.

The form, and consistency of the interactions with the underlying OSp$\left(
1|2\right) $ gauge symmetry, are determined by applying $X\cdot \partial $
or $X\cdot \gamma $ on both sides of (\ref{spin1/2}) and using the kinematic
equations in (\ref{spinhalf}) and (\ref{scalar}). This also produces the
conditions 
\begin{equation}
X^{2}A_{M}=0,\quad \left( X\cdot \partial +1\right) A_{M}=0,\quad X\cdot A=0
\end{equation}
on the gauge field. The same ``kinematic'' equations for the gauge field
also follow from other independent considerations, including consistency of
background fields with the Sp$\left( 2,R\right) $ gauge symmetry in the
two-time worldline formalism \cite{emgrav}, and the analysis in the
following section for higher spinning fields.

Finally, it is important to note that (\ref{spin1/2}) or (\ref{Lpsi}) have a
kappa type local fermionic symmetry given by 
\begin{equation}
\delta \Psi _{\alpha }=X_{M}\left( \gamma ^{M}\kappa \right) _{\alpha }
\end{equation}
where $\kappa _{\alpha }\left( X\right) $ is any spinor in $d+2$ dimensions.
To prove the kappa symmetry use $\gamma \cdot \left( \partial -iA\right) $ $%
\gamma \cdot X=-\gamma \cdot X\,\gamma \cdot \left( \partial -iA\right)
+X\cdot \left( \partial -iA\right) +\left( d+2\right) /2$ and apply the
kinematic conditions (\ref{spinhalf}). This means that only half of the
fermions are physical, in accord with what is expected when the two-time
theory is reduced from $d+2$ dimensions to $d$ dimensions.

In the case of free fields Dirac showed, with the parametrization given in (%
\ref{confpart}), that the solution space of these equations is precisely the
massless Dirac equation for a fermionic field in $d$ dimensions. This is
also the conclusion reached in \cite{spin} by quantizing the OSp$\left(
1|2\right) $ worldline theory in the gauge $X^{+^{\prime }}\left( \tau
\right) =1,$ $P^{+^{\prime }}\left( \tau \right) =0,$ $\psi ^{+^{\prime
}}\left( \tau \right) =0.$ To show how the $d$ dimensional theory is
embedded in $d+2$ we give here the field theory version of the gauge choice
used in the worldline approach. We first fix the kappa symmetry so that it
corresponds to the worldline fermionic gauge $\psi ^{+^{\prime }}\left( \tau
\right) =0$ 
\begin{equation}
\left( \gamma ^{+^{\prime }}\Psi \right) _{\alpha }=0.  \label{kappag}
\end{equation}
Then use the parametrization (\ref{confpart}) and the chain rule (\ref
{chain1},\ref{chain2}) to show that the operators applied on the gauge fixed 
$\Psi $ can be rewritten in the form 
\begin{eqnarray}
\gamma ^{M}\partial _{M}\Psi &=&\frac{1}{\kappa }\gamma ^{\mu }D_{\mu }\Psi +%
\frac{1}{\kappa }\left( \gamma ^{-^{\prime }}-x^{\mu }\gamma _{\mu }\right)
\Psi \\
\gamma ^{M}X_{M}\Psi &=&-\kappa \left( \gamma ^{-^{\prime }}-x^{\mu }\gamma
_{\mu }\right) \Psi
\end{eqnarray}
and 
\begin{eqnarray}
\gamma \cdot X\,\gamma \cdot \left( \partial -iA\right) \Psi &=&-\left(
\gamma ^{-^{\prime }}-x^{\mu }\gamma _{\mu }\right) \gamma ^{\mu }\left(
D_{\mu }-iA_{\mu }\right) \Psi \\
&&+\left( \gamma ^{-^{\prime }}-x^{\mu }\gamma _{\mu }\right) \left(
2\lambda -x^{2}\right) \left( \partial _{\lambda }-iA_{\lambda }\right) \Psi
\end{eqnarray}
where $D_{\mu }=\partial _{\mu }+x_{\mu }\partial _{\lambda }$ as in (\ref
{laplace}). Inserting these forms in the interacting equation of motion we
see that we remain with the overall factor $\left( \gamma ^{-^{\prime
}}-x^{\mu }\gamma _{\mu }\right) $ on both sides of the equation, but since
this is an ivertible matrix that satisfies $\left( \gamma ^{-^{\prime
}}-x^{\mu }\gamma _{\mu }\right) ^{2}=x^{2},$ it can be removed from both
sides. Furthermore by using $X^{2}=0$ we set $\lambda =x^{2}/2$ which
eliminates the last term in the last equation.

The result is the interacting massless Dirac field in $d$ dimensions, with SO%
$\left( d,2\right) $ conformal symmetry, in full agreement with the
worldline theory approach. Therefore, the content of (\ref{spinhalf},\ref
{spin1/2}) or (\ref{Lpsi}) using the $d$ dimensional coordinates (\ref
{confpart}) and kappa gauge (\ref{kappag}) is the interacting massless
fermionic field with SO$\left( d,2\right) $ conformal symmetry.

However, as discussed in \cite{spin} there are other gauge choices in
two-time physics which would lead to other physical interpretations for the
SO$\left( d,2\right) $ symmetry and of the dynamics from the point of view
of a one-time observer. Using the corresponding parametrization for $%
X^{M},\psi ^{M}$ we fully expect that the two-time field equations (\ref
{spinhalf}) would yield the same richness of $d$ dimensional spin 1/2
one-time physics, but now in the language of field theory.

\section{Vector and higher spin fields}

When $n=2,3,\cdots $ the fermions $\psi _{a}^{M}$ lead to higher spin
particles. To display the spin components of the wavefunction we adapt the
methods of \cite{town-etal} to the case of SO$\left( d,2\right) $. The $n$
anticommuting $\psi _{a}^{M}$ are represented in terms of SO$\left(
d+2\right) $ Dirac gamma matrices $\gamma _{\alpha \beta }^{M}$ acting in
spinor space labelled by $\alpha =1,\cdots ,2^{\left( d+2\right) /2}.$ They
are given in direct product form acting on the physical state with spin
components $|\Phi _{\alpha _{1}\alpha _{2}\cdots \alpha _{a}\cdots \alpha
_{n}}>$%
\begin{equation}
\psi _{a}^{M}=\gamma ^{\ast }\otimes \cdots \gamma ^{\ast }\otimes \frac{1}{%
\sqrt{2}}\gamma ^{M}\otimes 1\otimes \cdots 1
\end{equation}
where the $\frac{1}{\sqrt{2}}\gamma ^{M}$ is inserted in the $a$'th entry of
the direct product, and $\gamma ^{\ast }$ (analog of $\gamma _{5}$ in four
dimensions) is the product of all $d+2$ gamma matrices $\gamma ^{\ast
}=i^{\left( d+2\right) /2}\gamma ^{0^{\prime }}\gamma ^{1^{\prime }}\gamma
^{0}\cdots \gamma ^{d-1}$ such that $\left\{ \gamma ^{\ast },\gamma
^{M}\right\} =0$ and $\left( \gamma ^{\ast }\right) ^{2}=1$ (for simplicity,
we assume even $d+2\equiv 2r$. If $d+2$ is odd the spinor space $\alpha $ is
doubled to avoid $\gamma ^{\ast }$ proportional to identity).

In this formalism the constraint $\psi _{\lbrack a}\cdot \psi _{b]}=0$ (for $%
n\neq 0$) on the physical state is solved by the following spin
wavefunction. For even $n$ ($n\neq 2$) the spin wavefunction $<X|\Phi
_{\alpha _{1}\alpha _{2}\cdots \alpha _{a}\cdots \alpha _{n}}>$ is a bosonic
field written in terms of a SO$\left( d,2\right) $ tensor $F_{indices}\left(
X\right) $ whose indices correspond to a Young tableau shaped like a
rectangle, with $\left( d+2\right) /2$ columns and $n/2$ rows, as follows 
\begin{eqnarray}
\Phi _{\alpha _{1}\alpha _{2}\cdots \cdots \alpha _{n}} &=&\left( \gamma
^{M_{1}^{1}\cdots M_{\left( d+2\right) /2}^{1}}\right) _{\alpha _{1}\alpha
_{2}}\left( \gamma ^{M_{1}^{2}\cdots M_{\left( d+2\right) /2}^{2}}\right)
_{\alpha _{3}\alpha _{4}}\cdots \left( \gamma ^{M_{1}^{n/2}\cdots M_{\left(
d+2\right) /2}^{n/2}}\right) _{\alpha _{n-1}\alpha _{n}} \\
&&\times F_{[M_{1}^{1}\cdots M_{\left( d+2\right) /2}^{1}];[M_{1}^{2}\cdots
M_{\left( d+2\right) /2}^{2}];\cdots \lbrack M_{1}^{n/2}\cdots M_{\left(
d+2\right) /2}^{n/2}]}.  \nonumber
\end{eqnarray}
The indices on $F_{indices}$ have permutation properties associated with SO$%
\left( n\right) $ type Young tableaux : (i) the antisymmetric indices $%
[M_{1}^{i}\cdots M_{\left( d+2\right) /2}^{i}]$ correspond to the column $i$%
, (ii) the $n/2$ columns of indices for different $i$'s are symmetrized with
each other, (iii) under anti-symmetrization with one more index of a
neighboring column the wavefunction vanishes 
\begin{equation}
F_{[M_{1}^{1}\cdots M_{\left( d+2\right) /2}^{1};M_{1}^{2}]\cdots M_{\left(
d+2\right) /2}^{2}];\cdots \lbrack M_{1}^{n/2}\cdots M_{\left( d+2\right)
/2}^{n/2}]}=0,
\end{equation}
(iv) to insure irreducibility under SO$\left( d,2\right) $ a vanishing trace
for any pair of indices using $\eta ^{MN}$ is required, symbolically $%
F_{indices}\cdot \eta =0$.

For odd $n$ the spin wavefunction is a fermionic field $\psi
_{indices}^{\alpha }\left( X\right) $, whose indices correspond to the Young
tableau described above with $\left( n-1\right) /2$ columns, and there is
one leftover spinor index $\alpha ,$ which satisfies the irreducibility
condition 
\begin{equation}
\left( \gamma ^{M_{1}^{1}}\right) _{\alpha \beta }\psi _{\lbrack
M_{1}^{1}\cdots M_{\left( d+2\right) /2}^{1}];[M_{1}^{2}\cdots M_{\left(
d+2\right) /2}^{2}];\cdots }^{\beta }=0.
\end{equation}

For $n=2,$ there is an exception since $q\neq 0$ is possible for an SO$%
\left( 2\right) $ singlet. Then it is possible to get a singlet (gauge
invariant) of OSp$\left( 2|2\right) $ even though it is not necessarily
neutral under the subgroup SO$\left( 2\right) .$ This allows a more general
solution for $\Phi _{\alpha _{1}\alpha _{2}}$ than the above. Imposing $\psi
_{\lbrack 1}\cdot \psi _{2]}|\Phi >=2iq|\Phi >$ we find 
\begin{eqnarray}
\Phi _{\alpha _{1}\alpha _{2}}\left( X\right) &=&\left( 1+i\gamma ^{\ast
}sign\left( q\right) \,\left( -1\right) ^{p}\right) \left( \gamma
^{M_{1}\cdots M_{p+2}}\right) _{\alpha _{1}\alpha _{2}}F_{M_{1}\cdots
M_{p+2}}\left( X\right) ,  \label{n2F} \\
p &=&\frac{1}{2}\left( d-2\right) -\left| q\right| ={integer}.
\end{eqnarray}
Therefore, by adjusting the value of $q$ it is possible to obtain solutions
that correspond to antisymmetric tensors $F_{M_{1}\cdots M_{p+2}}\left(
X\right) $ with any of the values of $p$ in the set $\{-1,0,1,\cdots ,\frac{1%
}{2}\left( d-2\right) \}$. If $q=0$ only the last value of $p$ is possible.
This in contrast with the case of $n\geq 3$ for which only one solution is
possible as given above.

For the rest of the discussion, for simplicity, we will specialize to the $%
n=2$ case, and furthermore concentrate on free fields so we will relax the
conditions $X^{2}P^{2}\sim X\cdot \psi _{a}P\cdot \psi _{a}\sim 0$ of (\ref
{ppandp},\ref{psipsi}) to $P^{2}\sim P\cdot \psi _{a}\sim 0$ . The physical
state condition $\psi _{a}\cdot P\sim 0$ requires that the $n=2$
wavefunction $F_{M_{1}\cdots M_{p+2}}\left( X\right) $ given in (\ref{n2F})
is an on-shell field strength for a $p$-brane gauge potential $%
A_{M_{1}\cdots M_{p+1}}$%
\begin{equation}
F_{M_{1}\cdots M_{p+2}}\left( X\right) =\partial _{\lbrack
M_{p+2}}A_{M_{1}\cdots M_{p+1}]},\quad \partial ^{M_{p+2}}\partial _{\lbrack
M_{p+2}}A_{M_{1}\cdots M_{p+1}]}=0.  \label{W-eqs}
\end{equation}
The additional physical state condition in (\ref{ppandp}) requires a
specific dimension 
\begin{equation}
X\cdot \partial F_{M_{1}\cdots M_{p+2}}=-\left( \frac{d-2}{2}+2-\left|
q\right| \right) F_{M_{1}\cdots M_{p+2}}=-\left( p+2\right) \,F_{M_{1}\cdots
M_{p+2}}.  \label{Fdim}
\end{equation}
This equation holds provided the gauge field satisfies $\left( X\cdot
\partial +p+1\right) A_{M_{1}\cdots M_{p+1}}=\partial _{\lbrack
M_{1}}U_{M_{2}\cdots M_{p+1}]}$ for any $U_{M_{2}\cdots M_{p+1}}$. Through a
gauge transformation $\delta A_{M_{1}\cdots M_{p+1}}=\partial _{\lbrack
M_{1}}\Lambda _{M_{2}\cdots M_{p+1}]}$ one can eliminate $U$, hence $U$ is
arbitrary. With the choice $U_{M_{2}\cdots M_{p+1}}=X^{M_{1}}A_{M_{1}\cdots
M_{p+1}}$ the condition on $A_{M_{1}\cdots M_{p+1}}$ takes the gauge
invariant form $X^{M_{p+2}}F_{M_{1}\cdots M_{p+2}}=0.$

The last equation in (\ref{W-eqs}) may be modified to include interactions
through a conserved $p$ brane current, so the combined equations (\ref{W-eqs}%
,\ref{Fdim}) may be generalized to 
\begin{equation}
X^{M_{p+2}}F_{M_{1}\cdots M_{p+2}}=0,\quad \partial ^{M_{p+2}}F_{M_{1}\cdots
M_{p+2}}=J_{M_{1}\cdots M_{p+1}}.  \label{current}
\end{equation}
The last equation contracted with either $X^{M_{1}}$ or $\partial ^{M_{1}}$
shows that the brane current must be conserved, satisfy an additional
constraint, and be of definite dimension 
\begin{equation}
\partial ^{M_{1}}J_{M_{1}\cdots M_{p+1}}=0,\quad X^{M_{1}}J_{M_{1}\cdots
M_{p+1}}=0,\quad \left( X\cdot \partial +p+1\right) J_{M_{1}\cdots
M_{p+1}}=0.
\end{equation}
The first equation in (\ref{current}) is ``kinematics'' and the last is
dynamics. The dynamical equation follows from a varying the Lagrangian 
\begin{equation}
L_{d+2}^{A}=-\frac{1}{4}F_{M_{1}\cdots M_{p+2}}F^{M_{1}\cdots
M_{p+2}}+A^{M_{1}\cdots M_{p+1}}J_{M_{1}\cdots M_{p+1}}
\end{equation}
which has the gauge invariance for a $p+1$ gauge potential.

In the case of a vector potential we may identify it with the Yang-Mills
gauge potential that appeared in the previous section and which coupled to
the charged scalars or fermions. Then the current $J_{M}$ need not be
included as an additional source at it follows from the gauge couplings in $%
L_{d+2}^{\Phi }$ or $L_{d+2}^{\Psi }.$

The $p$ brane potential $A_{M_{1}\cdots M_{p+1}}$ satisfies similar
constraints to those of $J_{M_{1}\cdots M_{p+1}}$ after fixing some gauge
symmetries. Consider fixing the gauge $X^{M_{1}}A_{M_{1}\cdots M_{p+1}}=0.$
Then the first equation in (\ref{current}) reduces to $\left( X\cdot
\partial +p+1\right) A_{M_{1}\cdots M_{p+1}}=0,$ which requires $%
A_{M_{1}\cdots M_{p+1}}$ to have a definite dimension. Despite the gauge
choice there still remains gauge symmetry under $\delta A_{M_{1}\cdots
M_{p+1}}=\partial _{\lbrack M_{1}}\Lambda _{M_{2}\cdots M_{p+1}]},$ for all $%
\Lambda _{M_{2}\cdots M_{p+1}}$ that have dimension $p,$ i.e. $\left( X\cdot
\partial +p\right) \Lambda _{M_{2}\cdots M_{p+1}}=0.$ This is sufficient
gauge symmetry to further fix the gauge of $A_{M_{1}\cdots M_{p+1}}$ since
it now has a definite dimension. Thus, through these gauge choices we may
take a $A_{M_{1}\cdots M_{p+1}}$ that satisfies constraints similar to those
of the current 
\begin{equation}
X^{M_{1}}A_{M_{1}\cdots M_{p+1}}=0,\quad \partial ^{M_{1}}A_{M_{1}\cdots
M_{p+1}}=0,\quad \left( X\cdot \partial +p+1\right) A_{M_{1}\cdots
M_{p+1}}=0,
\end{equation}
while the dynamics simplifies to the gauge fixed form 
\begin{equation}
\partial ^{M}\partial _{M}A_{M_{1}\cdots M_{p+1}}=J_{M_{1}\cdots M_{p+1}}.
\end{equation}

If we specialize to $n=2$ and $p=0$ (or $\left| q\right| =\left( d-2\right)
/2$ ) the physical state is a vector gauge field $A_{M}$ that satisfies the
gauge invariant equations for $F_{MN}=\partial _{M}A_{N}-\partial _{N}A_{M}$ 
\begin{eqnarray}
X^{M}F_{MN} &=&0,\quad \partial ^{M}F_{MN}=J_{N},\quad \\
X^{M}J_{M} &=&0,\quad \partial ^{M}J_{M}=0, \\
\quad \left( X\cdot \partial +2\right) F_{MN} &=&0,\quad \left( X\cdot
\partial +1\right) J_{M}=0.
\end{eqnarray}
For the fixed gauge described above these equations become 
\begin{eqnarray}
\left( X\cdot \partial +1\right) A_{M} &=&0,\quad X^{M}A_{M}=0,\quad
\partial ^{M}A_{M}=0,\quad \partial ^{M}\partial _{M}A_{N}=J_{N}, \\
\left( X\cdot \partial +1\right) J_{M} &=&0,\quad X^{M}J_{M}=0,\quad
\partial ^{M}J_{M}=0.
\end{eqnarray}
These coincide with equations that appear in Dirac's paper \cite{Dirac} for
the vector gauge potential. They also are in agreement with the results of
the background field approach introduced in \cite{emgrav}.

Following Dirac, if we take $d+2=6$ dimensions, the solution of these field
equations in the parametrization of eq.(\ref{confpart}) is precisely
equivalent to Maxwell's equations for a gauge potential $A_{\mu }\left(
x\right) $ in $d=4$ dimensions identified as part of the six dimensional $%
A_{M}\left( X\right) $. The conformal symmetry of Maxwell's theory in four
dimensions is none other than the SO$\left( 4,2\right) $ Lorentz symmetry in
six dimensions.

As we have emphasized in the previous sections the parametrization of eq.(%
\ref{confpart}) is connected to one of the possible gauge choices in
two-time physics. Parametrizations that are related to other gauge choices
would reveal other physical content in the $d$ dimensional field theory.

\section{Gravity}

All of the interacting Lagrangians above can be coupled to gravity. To do so
we follow the prescription obtained in \cite{emgrav}. In the usual way we
need a metric $G_{MN}\left( X\right) $ or vielbein $E_{M}^{a}\left( X\right) 
$ and a spin connection for SO$\left( d,2\right) $ $\omega _{M}^{ab}\left(
X\right) $ in $d+2$ dimensions, but we also need an additional vector $%
V^{M}\left( X\right) $ constructed from a potential $W\left( X\right) .$
These fields satisfy the following kinematic equations 
\begin{equation}
\pounds _{V}G^{MN}=-2G^{MN},\quad V^{M}=\frac{1}{2}G^{MN}\partial
_{N}W,\quad G^{MN}\partial _{M}W{ }\partial _{N}W=4W  \label{homo}
\end{equation}
where $\pounds _{V}G^{MN}$ is the Lie derivative $\pounds _{V}G^{MN}=V\cdot
\partial G^{MN}-\partial _{K}V^{M}G^{KN}-\partial _{K}V^{N}G^{KM}.$
Furthermore the kinematic conditions we had earlier for the various fields
now take the form 
\begin{eqnarray}
\pounds _{V}\Phi &=&-\frac{d-2}{2}\Phi ,\quad \pounds _{V}\Psi _{\alpha }=-%
\frac{d}{2}\Psi _{\alpha },\quad V^{M}F_{MN}=0,  \label{kine} \\
W\left( X\right) \Phi &=&0,\quad W\left( X\right) \Phi =0,\quad W\left(
X\right) A_{M}=0.
\end{eqnarray}
where the ordinary derivatives in the Lie derivative $\pounds _{V}$ should
be replaced by covariant derivatives consistent with a local Lorentz
symmetry SO$\left( d,2\right) $ in tangent space. Thus, wherever there was
an explicit $X^{M}$ in flat space, it is now replaced by $V^{M}\left(
X\right) $ and wherever there was a Yang-Mills derivative $\partial
_{M}+A_{M}$ it is now promoted to a SO$\left( d,2\right) $ covariant
derivative $\partial _{M}+\omega _{M}+A_{M}$. Using these modifications the
Lagrangians $L_{d+2}^{\Phi },$ $L_{d+2}^{\Psi },$ $L_{d+2}^{A}$ constructed
earlier in this paper are generalized to couple to gravity consistently with
the underlying OSp$\left( n|2\right) $ gauge symmetries of two-time physics.
They should also be multiplied by a volume factor $\sqrt{G}=\det E$ that
satisfies $\pounds _{V}\sqrt{G}=\left( d+2\right) \sqrt{G}$ as it follows
from (\ref{homo}).

Next we would like to write down a Lagrangian $L_{d+2}^{G}$ for the
gravitational sector. But first we will deal with the kinematic constraints
in (\ref{homo}) by rewriting them in tangent space using the vielbein and
spin connection and giving them a more geometrical meaning. In particular
since the spin connection is a gauge field its field strength (the
curvature) must satisfy 
\begin{equation}
V^{M}R_{MN}^{ab}=0,\quad R_{MN}^{ab}=\partial _{M}\omega _{N}^{ab}-\partial
_{N}\omega _{M}^{ab}+\left[ \omega _{M},\omega _{N}\right] ^{ab}.
\end{equation}
like other gauge fields in (\ref{kine}). Similarly, the vielbein can also be
viewed as a gauge field. We define the covariant derivative of the vielbein
with respect to the spin connection 
\begin{equation}
D_{M}E_{N}^{a}=\partial _{M}E_{N}^{a}+\omega _{Mb}^{a}E_{N}^{b}.
\end{equation}
The torsion tensor is given by 
\begin{equation}
T_{MN}^{a}=D_{M}E_{N}^{a}-D_{N}E_{M}^{a}.
\end{equation}
So, we also take the torsion tensor to satisfy the transversality condition,
as a kinematic condition 
\begin{equation}
V^{M}T_{MN}^{a}=0.
\end{equation}
We define the Lie derivative of the vielbein by including the covariant
derivative using the spin connection; it may be rewritten in terms of the
torsion as follows 
\begin{eqnarray}
\pounds _{V}E_{M}^{a} &=&V^{N}D_{N}E_{M}^{a}+\partial _{M}V^{N}E_{N}^{a} \\
&=&V^{N}D_{[N}E_{M]}^{a}+V^{N}D_{M}E_{N}^{a}+\partial _{M}V^{N}E_{N}^{a} \\
&=&V^{N}T_{NM}^{a}+D_{M}V^{a} \\
&=&D_{M}V^{a}
\end{eqnarray}
we have used the transversality condition on the torsion and defined $%
V^{a}=E_{M}^{a}V^{M}$, and $D_{M}V^{a}=\partial _{M}V^{a}+\omega
_{M}^{ab}V_{b}$. If $\pounds _{V}E_{M}^{a}$ is contracted with another
vielbein we obtain $\pounds _{V}G_{MN}$ in the form 
\begin{equation}
\pounds _{V}G_{MN}=2\pounds _{V}E_{M}^{a}E_{Na}=2D_{M}V^{a}E_{Na}.
\end{equation}
Due to the condition (\ref{homo}) this quantity is $2G_{MN}.$ Multiplying
both sides by $E^{Na}$ we find 
\begin{equation}
\pounds _{V}E_{M}^{a}=E_{M}^{a},
\end{equation}
with 
\begin{equation}
E_{N}^{a}=D_{M}V^{a}=\partial _{M}V^{a}+\omega _{M}^{ab}V_{b}.
\end{equation}
This form has been previously suggested in \cite{vasilev}), we derived it
here from the homothety conditions (\ref{homo}) obtained in the worldline
formalism \cite{emgrav}.

Thus, the vielbein constructed in this way satisfies the kinematic condition
automatically while it is fully determined by the arbitrary functions $%
V^{a}\left( X\right) $ and $\omega _{M}^{ab}\left( X\right) $. The only
condition on the functions $V^{a},\omega _{M}^{ab}$ is that the curvature
and torsion be transverse to $V^{M}.$ Modulo this condition the vielbein,
and metric $G_{MN}$ are determined. This form solves automatically the first
equation in (\ref{homo}). Similarly, the last equation in (\ref{homo}) can
be rewritten in terms of $V^{a}$ 
\begin{equation}
W=V^{a}V_{a}.
\end{equation}
There remains the second equation in (\ref{homo}) that now takes the form 
\begin{equation}
V_{a}=\frac{1}{2}E_{a}^{M}D_{M}\left( V_{b}V^{b}\right) =E_{a}^{M}\left(
D_{M}V^{b}\right) V_{b}=E_{a}^{M}E_{M}^{b}V_{b}
\end{equation}
which is an identity since $E_{a}^{M}E_{M}^{b}=\delta _{a}^{b}.$ Thus, all
kinematic conditions for the gravitational field are fully solved by the
arbitrary functions $V^{a}\left( X\right) $ and $\omega _{M}^{ab}\left(
X\right) ,$ and the definition of metric through $E_{M}^{a}=D_{M}V^{a}.$
Note also that we can rewrite the torsion as follows 
\begin{equation}
T_{MN}^{a}=D_{[M}D_{N]}V^{a}=R_{MN}^{ab}V_{b},\quad or\quad
T_{cd}^{a}=R_{cd}^{ab}V_{b}.  \label{torsR}
\end{equation}
Therefore, the torsion is obtained from the curvature (the spin connection
included torsion).

From this construction we may deduce (using $\pounds _{V}E_{M}^{a}=E_{M}^{a}$
and $\pounds _{V}R_{MN}^{ab}=-2R_{MN}^{ab}$ as any other gauge field
strength) 
\begin{eqnarray}
\pounds _{V}V^{a} &=&V^{M}D_{M}V^{a}=V^{M}E_{M}^{a}=V^{a}, \\
\pounds _{V}R_{cd}^{ab} &=&\pounds _{V}\left(
E_{c}^{M}E_{d}^{N}R_{MN}^{ab}\right) =-4R_{cd}^{ab}, \\
\pounds _{V}T_{cd}^{a} &=&-3T_{cd}^{a}.
\end{eqnarray}
The transversality conditions $V^{M}R_{MN}^{ab}=0$ on the curvature and
torsion may be rewritten in tangent space 
\begin{equation}
V^{c}R_{cd}^{\,\,ab}=0,  \label{homo2}
\end{equation}
while $V^{c}T_{cd}^{\,\,a}=0$ is automatically satisfied thanks to (\ref
{torsR}). This is the only remaining kinematic condition on the
gravitational background as long as the primary building blocks are $V^{a}$
and $\omega _{M}^{ab}.$

We now turn our attention to the Lagrangian in the gravitational sector that
generates the dynamics for gravity (i.e. impose the analog of the Einstein
equations) in two-time physics. Naively the Lagrangian would be given by the
Riemann curvature scalar $R=R_{ab}^{ab}$ but we must seek a modification in
light of the constraints generated by $\pounds _{V}$ as in (\ref{homo}, \ref
{homo2}, \ref{kine}). Consistent coupling with these constraints requires
the form 
\begin{equation}
L_{d+2}^{G}\sim \left( \det E\right) \,R_{ab}^{ab}\,\Phi ^{2\left(
d-4\right) /d-2}
\end{equation}
where $\Phi $ is one (or a combination) of the scalar fields described
earlier. Typically the scalar that appears in the overall factors in the
Lagrangians we constructed up to now would be identified as the dilaton.

\section{Discussion}

Combining the Lagrangians for scalars, spinors, vectors and gravitons we
have a total Lagrangian that generates the dynamical equations of motion
through a variational principle, and couples all the fields to one another 
\begin{eqnarray}
L_{d+2} &=&L_{d+2}^{G}+L_{d+2}^{A}+L_{d+2}^{\psi }+L_{d+2}^{\phi } \\
L_{d+2}^{G} &\sim &\left( \det E\right) \,\Phi ^{2\left( d-4\right)
/d-2}R_{ab}^{ab}\, \\
L_{d+2}^{A} &=&-\frac{1}{4}\left( \det E\right) \,\Phi ^{2\left( d-4\right)
/d-2}F_{MN}F_{KL}G^{MK}G^{NL} \\
L_{d+2}^{\Psi } &=&\left( \det E\right) \left[ \bar{\Psi}\gamma \cdot
V\,\,\gamma \cdot \left( \partial -iA\right) \Psi -h\Phi ^{2/\left(
d-2\right) }\bar{\Psi}\gamma \cdot V\,\Psi +\bar{\Psi}\gamma \cdot V\,\xi %
\right] \\
L_{d+2}^{\Phi } &=&-\frac{1}{2}\Phi \partial _{M}\left( \sqrt{G}%
G^{MN}\partial _{N}\Phi \right) -\lambda \frac{\left( d-2\right) }{2d}\Phi
^{2d/\left( d-2\right) }\sqrt{G}.
\end{eqnarray}
The dynamical equations thus obtained must be supplemented with the
kinematic conditions in (\ref{kine}) and (\ref{homo2}). In the gravitational
sector $V^{a},\omega _{M}^{ab}$ are the primary fields, not $E_{M}^{a}$ or $%
G_{MN}.$

As discussed above Dirac's program for coming down from $d+2$ dimensions to $%
d$ dimensions can be implemented through many possible paths, ending up with
a choice of some $d$ dimensions embedded in $d+2$ dimensions. In this way
one arrives at different looking but non-trivially related interacting field
theories in $d$ dimensions. This is the new lesson learned from two-time
physics.

As mentioned earlier one may consider several scalars, spinors, vectors
etc., and build a $d+2$ Lagrangian that would reproduce the Standard Model
in one of the $d$=4 versions of a $d+2=6$ dimensional theory in two-time
physics. The natural choice of 4 dimensions among the 6 is the one given in (%
\ref{confpart}) since that is the one that corresponds to the massless
relativistic particle. It would be interesting to find out what one can
learn from the other choices of $4$ dimensions that would produce dual
versions of the Standard Model. In particular, can one discover
non-perturbative phenomena, relations among parameters, or new measurable
effects of the standard theory in particle physics? These questions remain
to be investigated in general as well as for the Standard Model itself.

The two-time formulation presented here and in \cite{conf}-\cite{emgrav} has
properties that touches upon other popular but little understood concepts in
the current literature. Among them duality, holography, AdS-CFT, background
independence are ideas that can be seen to be present in two-time physics is
some generalized sense. Holography can be compared to the fact that $d$
dimensions, which can be thought of as a surface around the bulk of $d+2$
dimensions, is sufficient to describe all of the physics contained in the
bulk. In our version of holography we go down two dimensions rather than one
and therefore there is not just one $d$ dimensional ``surface'' but many,
and this connects to our version of ``duality''. Duality can be compared to
the many versions of $d$ dimensional theories that are related and actually
represent the same $d+2$ dimensional theory (an analogy to M-theory). We
have already given a concrete example of the usual AdS-CFT correspondence at
the end of section (4.2), as seen from the point of view of two-time
physics, and this could be generalized to more interesting cases. Finally
concepts of background independence are present since one could start with a 
$d+2$ theory without backgrounds and end up with a theory in $d$ dimensions
with many possible curved backgrounds.

We need to end on a down note, but hopefully a stimulating one. The
formulation of two-time physics in field theory presented here is
incomplete. The fact that the subsidiary ``kinematic'' conditions are not
derived directly as an equation of motion from the field theory Lagrangian
is a sign that the formulation is incomplete. Surely one could introduce
Lagrange multipliers to impose these conditions, but this seems artificial.
Introducing a delta function $\delta \left( X^{2}\right) $ or $\delta \left(
V^{2}\right) $ in the action built from the Lagrangian above is also just as
artificial, and it still misses the other kinematic constraints due to 
\pounds $_{V}.$ Rather, a gauge principle that implements the underlying Sp$%
\left( 2,R\right) $ or OSp$\left( n|2\right) $ gauge symmetry directly in
field theory is the needed ingredient. This would generate all the kinematic
or dynamic constraints simultaneously, as it does in the worldline
formalism. In this sense the worldline formalism seems more fundamental at
the current stage.

Once the field theory formulation is completed it would then be possible to
investigate with more confidence second quantization in the formalism of
two-time physics, and try to establish the validity of the duality relations
among the d-dimensional theories at the second quantized level. Some such
duality is expected, but the correct ordering of operators (or corresponding
anomalies) may need further understanding.

To implement the Sp$\left( 2,R\right) $ gauge symmetry in field theory
suggested above it may be more natural to consider fields that are functions
over phase space $\Phi \left( X,P\right) .$ This appears to go in the
direction of non-commutative geometry, but with specific goals that are not
currently part of the thinking in non-commutative geometry. Perhaps it would
be helpful to investigate in this direction to complete the field theoretic
formulation of two-time physics.

\end{document}